\documentclass[runningheads]{llncs}
\usepackage{graphicx}
\usepackage{comment}
\usepackage{amsmath,amssymb} 
\usepackage{color}

\usepackage[pagebackref=true,breaklinks=true,letterpaper=true,colorlinks,bookmarks=false]{hyperref}
\usepackage{epsfig}
\usepackage{enumitem}
\usepackage{multirow}
\usepackage[percent]{overpic}
\usepackage{xcolor}
\usepackage[margin=0pt]{caption}
\usepackage[margin=0pt,labelformat=simple]{subcaption}

\usepackage{booktabs,dcolumn}
\usepackage{multirow}
\usepackage{array}
\newcolumntype{P}[1]{>{\centering\arraybackslash}p{#1}}
\newcolumntype{M}[1]{>{\centering\arraybackslash}m{#1}}
\usepackage{diagbox}
\usepackage{dblfloatfix}
\usepackage{breqn}
\usepackage{bm}
\usepackage{tabu}

\newcommand{\Tref}[1]{Table~\ref{#1}}

\newcommand{\fref}[1]{Fig.~\ref{#1}}

\newcommand{\sref}[1]{Sec.~\ref{#1}}

\newcommand{\eg}{e.g.~}
\newcommand{\ie}{i.e.~}
\renewcommand{\paragraph}[1]{\noindent\textbf{#1.}~}

\begin{document}
\pagestyle{headings}
\mainmatter
\def\ECCVSubNumber{807}  

\title{Learning the Loss Functions in \\ a Discriminative Space  for Video Restoration}

\titlerunning{Learning the Loss Functions in a Discriminative Space for Video Restoration}
%
\author{Younghyun Jo\inst{1}\and
Jaeyeon Kang\inst{1}\and
Seoung Wug Oh\inst{1}\and\\
Seonghyeon Nam\inst{1}\and
Peter Vajda\inst{2}\and
Seon Joo Kim\inst{1,2}}
%
\authorrunning{Y. Jo, J. Kang, S. W. Oh, S. Nam, P. Vajda, and S. J. Kim.}
%
\institute{\textsuperscript{1}Yonsei University, \textsuperscript{2}Facebook}
\maketitle

\begin{abstract}
With more advanced deep network architectures and learning schemes such as GANs, the performance of video restoration algorithms has greatly improved recently. 
Meanwhile, the loss functions for optimizing deep neural networks remain relatively unchanged.
To this end, we propose a new framework for building effective loss functions by learning a discriminative space specific to a video restoration task. 
Our framework is similar to GANs in that we iteratively train two networks -- a generator and a loss network.
The generator learns to restore videos in a supervised fashion, by following ground truth features through the feature matching in the discriminative space learned by the loss network.
In addition, we also introduce a new relation loss in order to maintain the temporal consistency in output videos.
Experiments on video super-resolution and deblurring show that our method generates visually more pleasing videos with better quantitative perceptual metric values than the other state-of-the-art methods. 
\end{abstract}

\section{Introduction}
The performance of video restoration such as video super-resolution and deblurring has been increasing by leaps and bounds recently by employing deep neural networks (DNNs).
More advanced DNN architectures \cite{Tao_2017_ICCV,sajjadi2018frame,jo2018deep,chu2018temporally,RBPN2019,wang2019edvr,yi2019progressive,hyun2017online,zhou2019spatio} and datasets \cite{su2017deep,nah2016deep,xue2019video,nah2019reds} have been introduced for video restoration, enabling significant progress in video restoration.  
In this paper, we focus on another key component of deep learning for video restoration -- loss functions. 

Most DNNs for video restoration are optimized to reduce a pixel loss (\eg L2 or L1), which is correlated with the peak signal-to-noise ratio (PSNR) -- a common measure for evaluating the image quality.
While the PSNR values keep improving with new deep architectures, there is still a large gap between the restoration results obtained by optimizing for the pixel loss and the human visual perception \cite{wang2004image,johnson2016perceptual,Zhang_2018_CVPR}. 
This is because measuring the average of pixel-wise image differences is limited in capturing perceptually relevant differences like high frequency details \cite{Ledig_2017_CVPR}.


To produce restoration outputs that are more pleasing for the human perception, the VGG \cite{simonyan2014very} based perceptual loss has emerged \cite{dosovitskiy2016generating,johnson2016perceptual}.
The perceptual loss is calculated on high-level image feature representations extracted from the pretrained VGG network, and it has shown to restore fine details better than the pixel loss.
However, the VGG network is trained for the image classification task, and it may not provide the most suitable feature space to measure the image difference for the restoration tasks. 
The perceptual loss sometimes produces undesired checkerboard artifacts when naively used \cite{dosovitskiy2016generating,johnson2016perceptual}.
In addition, it is difficult to incorporate temporal consistency as the perceptual loss processes each frame independently.

Another alternative for generating perceptually pleasing videos is adopting generative adversarial networks (GANs) \cite{goodfellow2014generative}.
In the GAN framework, the generator $G$ produces realistic results by learning to fool the discriminator $D$, which is trained to distinguish between fake and real samples. 
To preserve the original content for the restoration tasks and reduce excessively hallucinated high frequency details, the pixel loss and the perceptual loss are usually combined together with the adversarial loss.
This strategy has been adopted in many tasks including image-to-image translation \cite{isola2016image,zhu2017toward,chen2018sketchygan,wang2018pix2pixHD} and image super-resolution \cite{yu2016ultra,Ledig_2017_CVPR,Sajjadi_2017_ICCV,wang2018esrgan,zhang2019ranksrgan}.

\begin{figure}[t]
\small
\centering
\begin{subfigure}[b]{0.25\linewidth}
    \centering
    \includegraphics[width=\linewidth,trim={0cm 12.8cm 26cm 0cm},clip]{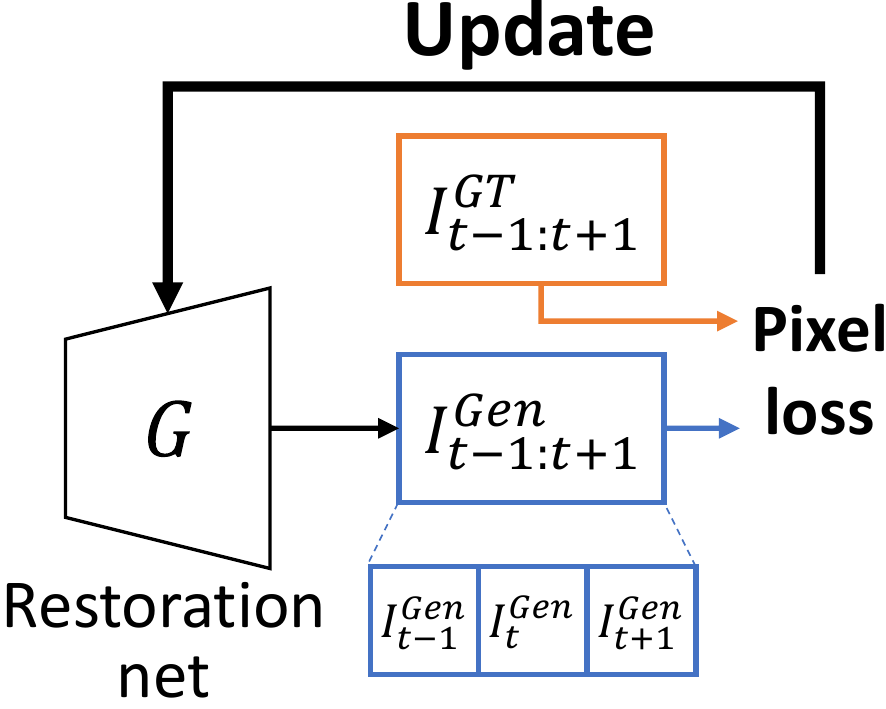}
    \caption{Pretrain G}
    \label{fig:teaser_a}
\end{subfigure}
\hfill
\begin{subfigure}[b]{0.33\linewidth}
    \centering
    \includegraphics[width=\linewidth,trim={9.6cm 13.5cm 13.8cm 0cm},clip]{figures/fig1}
    \caption{Initialize $T$}
    \label{fig:teaser_b}
\end{subfigure}
\begin{subfigure}[b]{0.38\linewidth}
    \centering
    \includegraphics[width=\linewidth,trim={21.6cm 12.4cm 0cm 0cm},clip]{figures/fig1}
    \caption{Alternately train $T$ and $G$}
    \label{fig:teaser_c}
\end{subfigure}
        
\caption{Our video restoration framework.
(a) We first pretrain a restoration network $G$ using the pixel loss.
(b) Then, the loss network $T$ is trained to discriminate between real videos and generated videos from $G$.
(c) $G$ and $T$ are updated alternately to improve other's performance. 
While the overall scheme is similar to training GANs, the generator in our framework is trained in a supervised manner using the feature space of $T$, not in an adversarial way.}
\label{fig:teaser}
\end{figure}
%
While GAN based video restoration tasks including video super-resolution \cite{xie2018tempogan,chu2018temporally} and video deblurring \cite{zhang2019adversarial} have been successfully used to generate realistic videos, there are still some major limitations.
First, the GAN based methods sometimes create excessive details or noise due to the adversarial loss, as $G$ is trained to deceive $D$ by hallucinating arbitrary details that may not be correlated with the corresponding ground truth data.
Second, training GANs is still challenging due to its instability.
To alleviate the difficulty in training, using modified objective functions \cite{arjovsky2017wasserstein,gulrajani2017improved,mao2017least,berthelot2017began} and regularization techniques \cite{salimans2016improved,miyato2018spectral} become essential.
Training GANs for videos is even more difficult as it also has to maintain the temporal consistency.
Additionally, training GANs involves extensive experiments as GANs are sensitive to network architectures and hyper-parameters.
In this paper, we propose a novel video restoration framework that outputs sharper videos compared to DNN methods that minimize the pixel loss, but without the excessive details of GAN based methods. 
We achieve this by employing a task specific loss network $T$, which is designed to be suitable for videos.
Unlike the fixed VGG network in the perceptual loss, our $T$ is optimized for a particular video restoration task.
The restoration network $G$ is updated using new loss functions computed from the learned feature space of $T$ in a supervised fashion, without the hassles of the adversarial loss (\fref{fig:teaser}).
In addition, we propose several mechanisms including the relation loss and a new fake sampling in order to improve the temporal consistency for video restorations. 
We apply our framework for video super-resolution and deblurring in this work.

Our contributions can be summarized as follows:
\begin{itemize}[noitemsep,topsep=0pt,leftmargin=*]
    \item We propose new loss functions for video restoration tasks by learning the discriminative feature space ($T$) to compare the generated videos and the ground truth videos.
    This enables the direct supervision for training $G$ without the adversarial loss,
    making our method free from common issues in GAN-based methods -- training instability and excessive detail hallucination.
    
    \item Compared to using the VGG network trained for the image classification for the perceptual loss, our loss network $T$ provides an effective perceptual loss optimized for a specific video restoration task. 
    
    \item We also propose a novel relation matching loss, designed to maintain the temporal consistency between frames.  
    In addition, a new way to construct fake samples that improve the discriminating capability of the loss network is presented. 
    
    \item We provide a general framework for the video restoration problems, therefore our method can be paired with any type of generator networks.
    
    
\end{itemize}

\section{Related work}
%
The pixel loss based video super-resolution methods have focused on designing effective network architectures to improve the performance \cite{kappeler2016video,Caballero_2017_CVPR,Tao_2017_ICCV,sajjadi2018frame,jo2018deep,RBPN2019,wang2019edvr,yi2019progressive}.
Similarly, many previous video deblurring works have investigated different architecture designs to enhance the results \cite{su2017deep,hyun2017online,Nah_2019_CVPR,wang2019edvr,zhou2019spatio}.
While the works that are optimized over the pixel loss have improved the performance in terms of PSNR, the perceptual quality has not improved as much as PSNR.
This is because the mean-square-error based evaluation is limited in capturing the difference in high frequency details that are essential for the perceptual quality.   

To enhance the perceptual quality for video restoration, a perceptual loss based method \cite{zhou2019spatio} and GAN based methods \cite{bhattacharjee2017temporal,xie2018tempogan,chu2018temporally,zhang2019adversarial} for video restoration have emerged.
In some of those works, several ways to maintain the temporal consistency between output frames have been introduced. 
In \cite{xie2018tempogan}, two discriminators were used -- one for the spatial sharpness and the other for the temporal smoothness. 
The method additionally enforces temporal consistency by using the optical flow. 
A spatio-temporally fused discriminator and the ping-pong loss were proposed in \cite{chu2018temporally} to generate realistic and temporally coherent video super-resolution results. 
The method also used optical flow for warping frames. 
While these attempts have definitely helped to maintain the temporal consistency, the methods can suffer from errors in the optical flow estimation. 
In \cite{bhattacharjee2017temporal}, local and global similarity losses were proposed to maintain the temporal consistency in video prediction.
However, the method assumes slowly varying motion between frames, which can be violated often in practice. 
In comparison, we propose to use a relation matching loss to maintain the temporal relationship between frames without using the optical flows or assuming slow motions.

Another closely related works to ours are the methods that use the discriminator feature matching loss 
\cite{li2016deep,wang2018vid2vid,chu2018temporally,wang2018perceptual}.
Similar to our work, a loss measured in the feature space learned in the discriminator was used to assist the adversarial loss for the stability. 
The big difference is that we completely remove the adversarial loss for training the generator, and the video restoration depends on the trained loss network in a supervised fashion in our work.
This makes training our method much easier and stable, as we do not have to deal with the hassles of training GANs.
More detailed description will be given in \sref{subsec:g_loss}.

The perceptual loss \cite{dosovitskiy2016generating,johnson2016perceptual} using the VGG \cite{simonyan2014very} features has also been used  to enhance the perceptual quality of restored images and videos \cite{wang2018vid2vid,chu2018temporally,zhang2019adversarial,zhou2019spatio}. 
However, the VGG network is trained for the image classification, so it may not provide the best feature space for the loss function in the video restoration tasks.
In comparison, our loss function uses the feature space of a loss network that is specifically trained for the video restoration task at hand. 

\section{Method}
\label{sec:method}
\begin{figure}[t]
\small
\centering
    \includegraphics[width=0.6\linewidth,trim={0cm 10.3cm 13.4cm 0cm},clip]{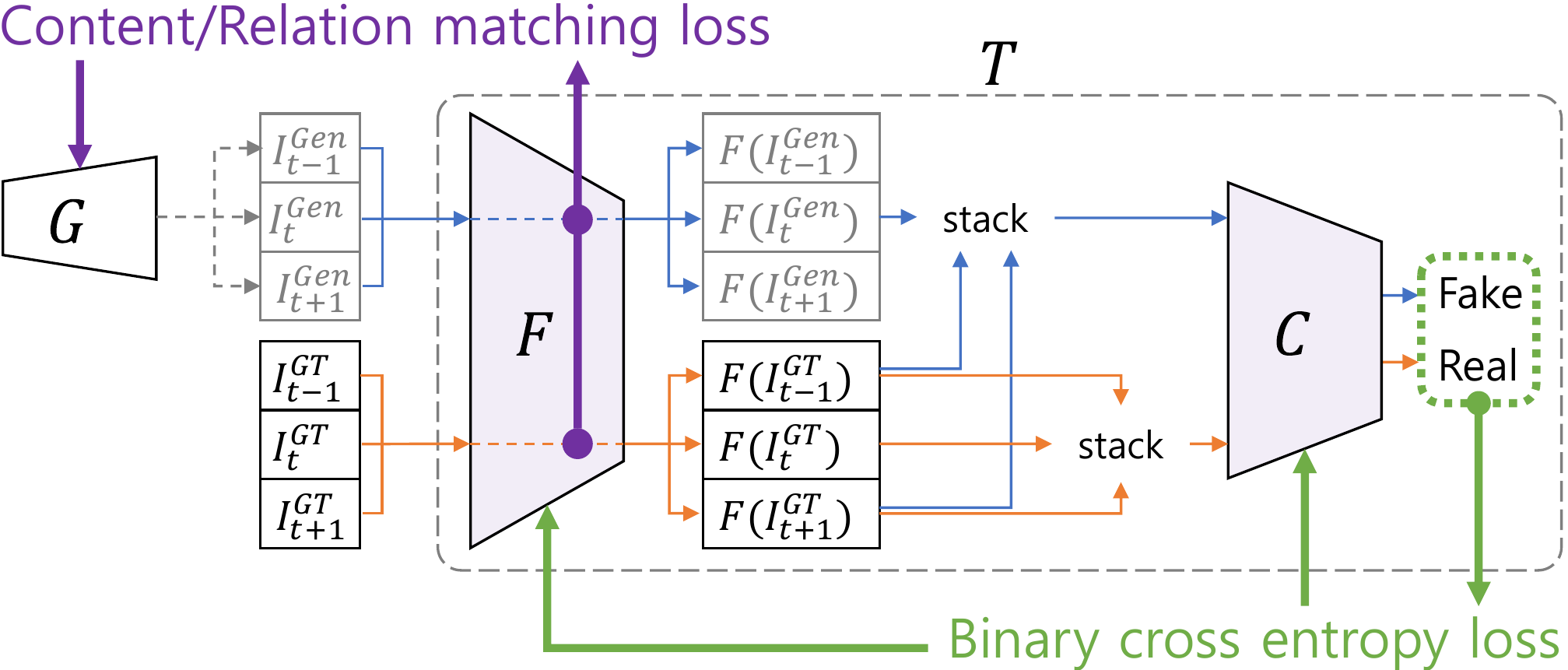}
\caption{Our framework consists of a restoration network $G$ and a loss network $T$.
$T$ is divided into a feature extractor $F$ and a sequence classifier $C$, and it
is trained through the binary cross entropy loss to distinguish between real and fake sequences.
$G$ is trained in a supervised fashion by the ground truth data through the proposed losses on the multi-scale feature maps of $F$.}
\label{fig:overview}
\end{figure}
Our method consists of a generator $G$ and a loss network $T$ (\fref{fig:overview}).
$T$ learns a new feature space for defining the loss functions for $G$ in the process of training to discriminate generated video sequences from real sequences.
$G$ is trained by our content and relation matching loss functions in addition to the pixel loss, and performs the video restoration task after the training.
We note that $I^{G\!T}$ is the ground truth frames and $I^{G\!e\!n}$ is the output frames of $G$.

\subsection{Loss Network ($T$)} 
In general, processing videos with CNNs can be implemented with 
3D convolutional layers to maintain the time axis for multiple input frames, or 2D convolutional layers for a multi-frame input concatenated channel-wise.
We empirically found that maintaining each frame's features in early convolutional layers and later combining their information works better for $T$ in our work.
This is similar to the late fusion strategy in \cite{karpathy2014large}.
With this strategy, we design $T$ that consists of a feature extractor $F$ and a multi-frame classifier $C$ as shown in \fref{fig:overview}.
Detailed network configurations for $F$ and $C$ are shown in \Tref{tab:d_structure}.

\begin{table}[t]
\small  
\centering
\vspace{-6pt}
\caption{Network structures of $F$ and $C$.
$F$ keeps all the input frame's features, and they are combined in $C$ for the binary sequence classification.}
\label{tab:d_structure}
\begin{subtable}[h]{0.52\linewidth}
    \centering
    \caption{Network structure of $F$.}
    \label{tab:f_structure}
    \begin{tabular}
        {@{}p{0.42\linewidth}@{\hskip1pt}p{0.22\linewidth}@{\hskip1pt}p{0.18\linewidth}@{\hskip1pt}p{0.14\linewidth}@{}}
        Layers & Kernel & Stride & Filters \\
        \hline
        conv3d-bn3d & $1\times7\times7$ & $(1,2,2)$ & 64 \\
        lrelu-conv3d-bn3d & $1\times3\times3$ & $(1,2,2)$ & 128 \\
        lrelu-conv3d-bn3d & $1\times3\times3$ & $(1,2,2)$ & 256 \\
        lrelu-conv3d-bn3d & $1\times3\times3$ & $(1,2,2)$ & 512 \\
    \end{tabular}
\end{subtable}
\hfill
\begin{subtable}[h]{0.46\linewidth}
    \centering
    \caption{Network structure of $C$.}
    \label{tab:c_structure}
    \begin{tabular}
        {@{}p{0.34\linewidth}@{\hskip1pt}p{0.25\linewidth}@{\hskip1pt}p{0.20\linewidth}@{\hskip1pt}p{0.17\linewidth}@{}}
        Layers & Kernel & Stride & Filters \\
        \hline
        lrelu-conv3d & $3\times3\times3$ & $(1,2,2)$ & 256 \\
        lrelu-conv3d & $1\times1\times1$ & $(1,1,1)$ & 1 \\
    \end{tabular}
\end{subtable}
\end{table}
\paragraph{Feature extractor}
$F$ takes an image sequence $(I_{t-1}^{G\!e\!n}, I_{t}^{G\!e\!n}, I_{t+1}^{G\!e\!n})$, and outputs feature maps of each frame $F(I_{t-1}^{G\!e\!n})$, $F(I_{t}^{G\!e\!n})$, and $F(I_{t+1}^{G\!e\!n})$.
Likewise, for a ground truth sequence $(I_{t-1}^{G\!T}, I_{t}^{G\!T}, I_{t+1}^{G\!T})$, it outputs $F(I_{t-1}^{G\!T})$, $F(I_{t}^{G\!T})$, and $F(I_{t+1}^{G\!T})$.
$F$ consists of four convolutional layers (\ie $J=4$), each followed by a batch normalization layer and a leaky relu layer.
The negative slope of all leaky relu activations is set to 0.1.
Convolutional layers in $F$ have $1\times3\times3$ kernels except for the first layer which has $1\times7\times7$ kernels, and the stride size of $(1,2,2)$ is used to make the output size by $\frac{1}{2}$ at each convolutional layer.
Note that the kernel and the stride values are in the order of time, height, and width respectively.
The convolutional kernels in $F$ are not applied to the time axis, but the normalization layers use the statistics of the feature maps including the temporal information.
We have found that this design choice helps $G$ to produce better results.

\paragraph{Sequence classifier}
A new fake feature sequence $(F(I_{t-1}^{G\!T}), F(I_{t}^{G\!e\!n}), F(I_{t+1}^{G\!T}))$ and a real feature sequence $(F(I_{t-1}^{G\!T}), F(I_{t}^{G\!T}), F(I_{t+1}^{G\!T}))$ are fed into the classifier $C$, which consists of two convolutional layers. 
$C$ determines whether the input sequence is fake or real.
To make a fully convolutional framework, $C$ has PatchGAN \cite{isola2016image} discriminator structure.
In our framework, the ground truth data for the neighboring frames are used to form the new fake sample.
Note that this is different from the previous methods, where the fake sequence is constructed with all the generated frames (\eg $(F(I_{t-1}^{G\!e\!n}), F(I_{t}^{G\!e\!n}), F(I_{t+1}^{G\!e\!n}))$).
The new fake sample setting has only one fake frame information, so $F$ needs to extract more elaborated features to classify it as fake in $C$.
Since our proposed loss functions for updating $G$ are defined by the feature space of $F$, $G$ can produce better results as $F$ learns more advanced features.


With our approach, we can express the training formulation for $F$ and $C$ as follows:
\begin{equation}
    \begin{aligned}
        \mathcal{L}_{F,C} = &-\log C(F(I^{G\!T}_{t-1}), F(I^{G\!T}_{t}), F(I^{G\!T}_{t+1})) \\
        &-\log(1-C(F(I^{G\!T}_{t-1}), F(I^{G\!e\!n}_{t}), F(I^{G\!T}_{t+1}))).
    \end{aligned}
    \label{eq:d_loss}
\end{equation}

\subsection{Restoration Network ($G$)}
\label{subsec:g_loss}
While many types of GANs have shown impressive results for different tasks, training GANs is still a difficult problem.
Ideally, a well trained GAN will reach the equilibrium in which the generator will match the underlying data distribution. 
This rarely happens in practice as it is difficult to make the generator to fool the discriminator perfectly \cite{goodfellow2014generative,goodfellow2016nips,Martin2017TOWARDS}.

In addition, while the adversarial loss can move the generated samples toward the given data distribution, it does not guarantee that the generated samples match the corresponding ground truth data.
To deal with this issue, several reconstruction losses including the pixel loss and the VGG based perceptual loss have been used together to preserve the contents.
However, the common reconstruction losses fall short in preventing arbitrary details generated by using the adversarial loss.

\begin{figure}[t]
\centering
\begin{subfigure}[b]{0.28\linewidth}
    \centering
    \includegraphics[width=\linewidth,trim={0cm 0cm 28.6cm 15.6cm},clip]{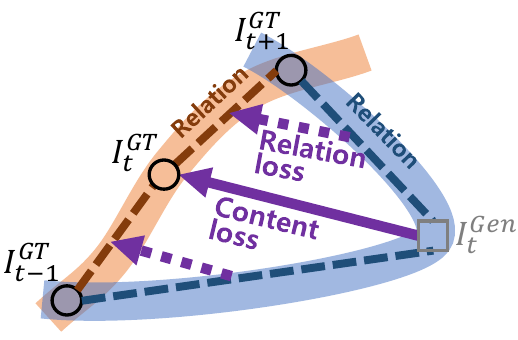}
    \caption{Concept.}
    \label{fig:loss_overview}
\end{subfigure}
\hfill
\begin{subfigure}[b]{0.66\linewidth}
    \centering
    \includegraphics[width=0.30\linewidth,trim={0cm 11.4cm 26.4cm 0cm},clip]{figures/fig3}
    \hfill
    \includegraphics[width=0.64\linewidth,trim={18.2cm 11.4cm 0cm 0cm},clip]{figures/fig3}
    \caption{Content and relation matching losses.}
    \label{fig:content_loss}
\end{subfigure}
\caption{Two proposed training objectives for updating $G$.
(a) The content matching loss is designed to preserve details in each frame, and the relation matching loss is for maintaining the temporal consistency.
(b) Content matching loss is calculated on multi-scale feature space of $F$ between $I^{G\!e\!n}_{t}$ and $I^{G\!T}_{t}$. 
To ensure the temporal consistency, we additionally calculate the relation matching loss between frames.
The relationship between $I^{G\!T}_{t-1}$ and $I^{G\!e\!n}_{t}$, which is represented on the distance embedded by $\phi$, is trained to be close to ground truth relationship between $I^{G\!T}_{t-1}$ and $I^{G\!T}_{t}$.
The same thing is done for the other time direction such as between the time $t$ and $t+1$.}
\label{fig:proposed_losses}
\end{figure}

To this end, we train $G$ in a supervised manner as the ground truth frames are given in the video restoration tasks.
In our framework, $G$ still utilizes $T$ by taking advantage of the features learned in $T$ for defining the loss functions.
The intuition behind this is that a discriminator $D$ ($T$ in our framework) converges fast in practice and reaches its near optimal \cite{goodfellow2014generative,goodfellow2016nips,Martin2017TOWARDS}.
So, instead of trying to train $G$ to beat $D$ in an unsupervised manner as in GAN, we aim to give a direct supervision for updating $G$ by matching the features of a generated frame $F(I^{G\!e\!n}_{t})$ to those of corresponding ground truth frame $F(I^{G\!T}_{t})$ using the following two proposed loss functions (\fref{fig:proposed_losses}).

\paragraph{Content matching loss}
First we propose the content matching loss $\mathcal{L}_C$, which makes $I^{G\!e\!n}_{t}$ to be close to $I^{G\!T}_{t}$ by reducing the distance on the feature space of $F$.
All the feature maps of $F$ are used for multi-scale matching.
We can formulate this as follows:
\begin{equation}
    \begin{aligned}
        \mathcal{L}_{C}^{j} = \mathcal{H}(F^{j}(I^{G\!e\!n}_{t}), F^{j}(I^{G\!T}_{t})), 
    \end{aligned}
    \label{eq:content_loss}
\end{equation}
where $j$ denotes the layer index of $F$, and $\mathcal{H}$ denotes the Huber loss.
The Huber loss is expressed as follows:
\begin{equation}	
    \mathcal{H}(\hat{Y}, Y) = \left\{
    \begin{array}{ll}
    \frac{1}{2}|| \hat{Y} - Y ||_2^2 & || \hat{Y} - Y ||_1 \leq \delta, \\
    \delta || \hat{Y} - Y ||_1 - \frac{1}{2}\delta^2  & \text{otherwise.} \\
    \end{array} 
    \right.  \\
\end{equation}

The proposed content matching loss function has the same form as the feature matching regularization technique introduced in \cite{salimans2016improved}, which is expressed as $||\mathbb{E}_{\bm{x}\sim p_{d\!a\!t\!a}}f(\bm{x})-\mathbb{E}_{\bm{z}\sim p_{\bm{z}}}f(G(\bm{z}))||^2_2$, where $f$ denotes the activations on an intermediate layer of a discriminator.
This is used as an auxiliary function for the regular adversarial loss to help the distribution matching in the unsupervised learning.
In contrast, we get away from using the adversarial loss entirely, replacing the adversarial loss with our content loss for the supervised learning. 

%
The discriminator feature matching loss \cite{li2016deep,wang2018vid2vid,chu2018temporally,wang2018perceptual}, which has been suggested to assist the adversarial loss for the stability, also has the same form as our content matching loss.
Once again, we no longer use the adversarial loss for training the generator, and our content matching loss is motivated and designed for a different purpose.
With our content matching loss, $G$ is directly guided only by the ground truth data in the feature space of $F$ rather than being guided by the unsupervised signal from the adversarial training.
Our method can give more robust signals to update $G$, therefore $G$ produces better results that are closer to the ground truth data instead of generating arbitrary details that come from the adversarial loss.
Additionally, we can avoid the training instability problems in the adversarial training that include saturated gradients \cite{mao2017least} and unstable gradients \cite{Martin2017TOWARDS}.

The content matching loss is also related to the perceptual loss \cite{dosovitskiy2016generating,johnson2016perceptual}.
The perceptual loss is expressed as $||g_{j}(\hat{y})-g_{j}(y)||^2_2$, where $g_{j}$ denotes the activation of the $j$-th layer of the VGG network \cite{simonyan2014very}, $y$ is the ground truth target image, and $\hat{y}$ is the generated image.
Most GAN based restoration methods have additionally used this loss for preserving the contents, but the VGG network is trained for the image classification task, so it may not provide adequate gradients for the video restoration task.
In comparison, as our $T$ is trained for a specific restoration task, it is more reasonable to use our $T$ as the perceptual loss network.
In this regard, one can consider learning $T$ as learning a loss function for a particular video restoration task, and our $T$ can provide more helpful gradients for updating $G$ over time.
Therefore, our $G$ produces more accurate and visually pleasing results than using the VGG based perceptual loss, and this will be shown in the experiments.
In addition, the VGG network does not need to be loaded to GPU memory, which saves the memory space and helps to shorten the training time.
This enhances the practical use of our algorithm as video data require much more memory.

\paragraph{Relation matching loss}
When we move $I^{G\!e\!n}_{t}$ to $I^{G\!T}_{t}$, we also have to consider the relationship with surrounding frames as maintaining the temporal consistency is important in video restoration tasks. 
Therefore, we suggest another cost function called the relation matching loss, which makes the temporal difference between $I^{G\!e\!n}_{t}$ and its surrounding ground truth frames to be close to those of a real sequence. 
We can formulate the relation matching loss as follows:
\begin{equation}
    \begin{aligned}
        \mathcal{L}^{j}_{R,n} = \mathcal{H}(\phi^{G\!e\!n,j}_{t+n}, \phi^{G\!T,j}_{t+n}), \\
    \end{aligned}
    \label{eq:relation_loss}
\end{equation}
where the function $\phi$ measures the distance between two features in different time in the Huber space as $\phi^{\psi,j}_{t+n} = \mathcal{H}(F^j(I^{\psi}_t),F^j(I^{G\!T}_{t+n}))$, and $n$ is the time difference from the reference time $t$.
The term $\phi^{G\!T,j}_{t+n}$ serves as the ground truth relation, and the loss function makes the generated relation $\phi^{G\!e\!n,j}_{t+n}$ get close to the ground truth relation.

Instead of simply extending an image discriminator to the time axis for videos, we introduce the relation matching loss to force the temporal consistency more effectively.
As we treat supervised tasks and the output is deterministic, we can use $(I^{G\!T}_{t-1}, I^{G\!e\!n}_{t})$ pair instead of $(I^{G\!e\!n}_{t-1}, I^{G\!e\!n}_{t})$ in $\mathcal{L}_R$ for updating the generator with more clear signal.
The input $I^{G\!e\!n}_{t}$ can be guided by the ground truth $I^{G\!T}_{t-1}$ in the former case, while the training signal is less clear in the latter case as both inputs are non-stationary.
This loss function makes transitions between frames more smooth and prevents excessively sharpened results.
It is also useful as it does not have to assume slowly varying motion, and does not require assistance from optical flow as in the previous methods.

\paragraph{Pixel loss}
We additionally define a loss in the pixel space to prevent color permutation as follows:
\begin{equation}
\begin{aligned}
    \mathcal{L}_P = \mathcal{H}(I^{G\!e\!n}_{t}, I^{G\!T}_{t}). \\
\end{aligned}
\label{eq:pixel_loss}
\end{equation}

\paragraph{Total loss}
From the proposed loss functions, total loss for updating $G$ is as follows:
\begin{equation}
\begin{aligned}
    \mathcal{L}_G = \frac{1}{J} \sum_{j=1}^{J} &\mathcal{L}_{C}^{j} + \frac{1}{2NJ} \sum_{n=-N,n\neq0}^{N} \sum_{j=1}^{J} \mathcal{L}_{R,n}^{j} + \mathcal{L}_{P}, \\
\end{aligned}
\label{eq:total_loss}
\end{equation}
where $J$ denotes the number of layers in $F$, $N$ denotes the temporal radius.
In our method, we set $N=1$ and all thresholds of Huber loss as 0.01.

For training, we first train $T$ only for a small number of iterations for the initialization.
Then, we alternately train $T$ and $G$ similar to the GAN training. 
As the training progresses, $T$ becomes better at distinguishing between the generated and the real samples, and $G$ produces more accurate outputs. 
Note that we can use any of the previous architectures for $G$ by simply replacing the loss function.

One crucial advantage of our method is that we do not need to run extensive experiments to find the balancing weights between the reconstruction loss and the adversarial loss. 
In general, the weight of the adversarial loss determines how much details to create, and much effort is needed to find a proper value.
However, we found that our loss function is less affected by the weight balancing because all the loss terms use the ground truth data as the reference (see the supplementary materials for details). 

\section{Experiments}
\label{sec:experiments}
We show the effectiveness of our method on two video restoration tasks: video super-resolution and video deblurring.
Four metrics are used to quantitatively evaluate the quality of results.
The first metric is the learned perceptual image patch similarity (LPIPS) \cite{Zhang_2018_CVPR}, which measures the perceptual similarity using deep features.
LPIPS is designed to measure the quality of images from the perspective of human visual perception, and we used LPIPS version $0.1$.
To examine the temporal consistency, we adopt \textit{tOF} measure introduced in \cite{chu2018temporally} to compute the motion similarity between the result and the ground truth as $||OF(I^{G\!e\!n}_{t-1}, I^{G\!e\!n}_{t}) - OF(I^{G\!T}_{t-1}, I^{G\!T}_{t})||$, and we set $OF$ as Gunnar Farneback's optical flow algorithm \cite{farneback2003two}.
We also evaluate the peak signal to noise ratio (PSNR) and the structural similarity index (SSIM) \cite{wang2004image}, which are traditionally used for image quality assessment.
Lower values are better for LPIPS and tOF, and higher values are better for PSNR and SSIM.
PSNR and SSIM are known to have difficulties in accounting for many nuances of human visual perception \cite{Zhang_2018_CVPR}, and we consider LPIPS and tOF to be more important in our work.

\subsection{Video super-resolution}
\begin{figure}[t]
\centering
\begin{subfigure}[h]{\linewidth}
    \centering
    \begin{tabular}
        {@{}p{0.12\linewidth}@{\hskip2pt}}
        ~\\
        \centering\scriptsize~Foliage/10 \\
        \includegraphics[width=\linewidth]{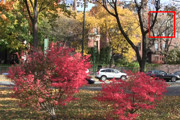} \\
        ~\\
        ~\\
        ~\\
        ~\\
        ~\\
        \centering\scriptsize~\mbox{Building/11} \\
        \includegraphics[width=\linewidth]{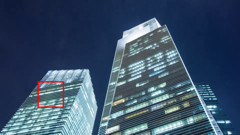} \\
    \end{tabular}%
    \begin{tabular}
        {@{}p{0.14\linewidth}@{\hskip2pt}p{0.14\linewidth}@{\hskip2pt}p{0.14\linewidth}@{\hskip2pt}p{0.14\linewidth}@{\hskip2pt}p{0.14\linewidth}@{\hskip2pt}p{0.14\linewidth}@{}}
        \centering\scriptsize~\mbox{EnhanceNet} &
        \centering\scriptsize~\mbox{EDVR} & 
        \centering\scriptsize~\mbox{Ours(EDVR)} & 
        \centering\scriptsize~\mbox{TecoGAN} & 
        \centering\scriptsize~\mbox{Ours(TecoGAN)} & 
        \centering\scriptsize~\mbox{GT} \tabularnewline
        
        \includegraphics[width=\linewidth]{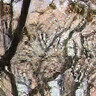} &
        \includegraphics[width=\linewidth]{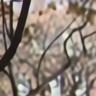} &
        \includegraphics[width=\linewidth]{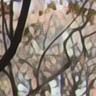} &
        \includegraphics[width=\linewidth]{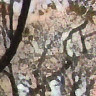} &
        \includegraphics[width=\linewidth]{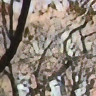} &
        \includegraphics[width=\linewidth]{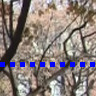} \tabularnewline
        
        \includegraphics[width=\linewidth]{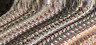} &
        \includegraphics[width=\linewidth]{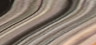} &
        \includegraphics[width=\linewidth]{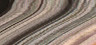} &
        \includegraphics[width=\linewidth]{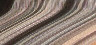} &
        \includegraphics[width=\linewidth]{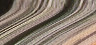} &
        \includegraphics[width=\linewidth]{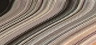} \tabularnewline
        
        \includegraphics[width=\linewidth]{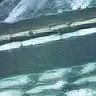} &
        \includegraphics[width=\linewidth]{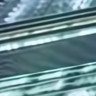} &
        \includegraphics[width=\linewidth]{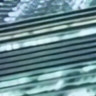} &
        \includegraphics[width=\linewidth]{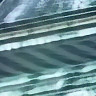} &
        \includegraphics[width=\linewidth]{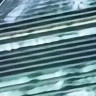} &
        \includegraphics[width=\linewidth]{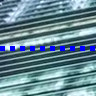} \tabularnewline
        
        \includegraphics[width=\linewidth]{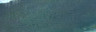} &
        \includegraphics[width=\linewidth]{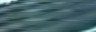} &
        \includegraphics[width=\linewidth]{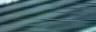} &
        \includegraphics[width=\linewidth]{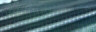} &
        \includegraphics[width=\linewidth]{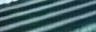} &
        \includegraphics[width=\linewidth]{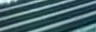} \tabularnewline

    \end{tabular}%
    \caption{Qualitative results on Foliage (Vid4) and Building (non-Vid4) scenes.}
    \label{fig:vsr_qualitative_vid4}
    \vspace{6pt}
\end{subfigure}
\begin{subfigure}[h]{\linewidth}
    \centering
    \begin{tabular}
        {@{}p{0.12\linewidth}@{\hskip2pt}}
        ~\\
        \centering\scriptsize 000/79 \\
        \includegraphics[width=\linewidth]{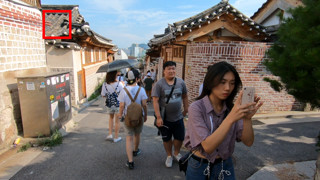} \\
    \end{tabular}%
    \begin{tabular}
        {@{}p{0.12\linewidth}@{\hskip2pt}p{0.12\linewidth}@{\hskip2pt}p{0.12\linewidth}@{}}
        \centering\scriptsize~\mbox{EDVR} &
        \centering\scriptsize~\mbox{Ours(EDVR)} &
        \centering\scriptsize~\mbox{GT} \tabularnewline
        
        \includegraphics[width=\linewidth]{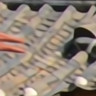} &
        \includegraphics[width=\linewidth]{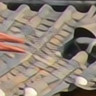} &
        \includegraphics[width=\linewidth]{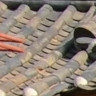} \tabularnewline
    \end{tabular}%
    \begin{tabular}
        {@{}@{\hskip2pt}p{0.12\linewidth}@{\hskip2pt}}
        ~\\
        \centering\scriptsize 004/10 \\
        \includegraphics[width=\linewidth]{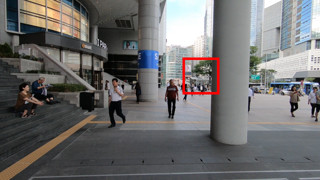} \\
    \end{tabular}%
    \begin{tabular}
        {@{}p{0.12\linewidth}@{\hskip2pt}p{0.12\linewidth}@{\hskip2pt}p{0.12\linewidth}@{}}
        \centering\scriptsize~\mbox{EDVR} &
        \centering\scriptsize~\mbox{Ours(EDVR)} &
        \centering\scriptsize~\mbox{GT} \tabularnewline
        
        \includegraphics[width=\linewidth]{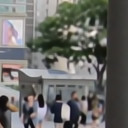} &
        \includegraphics[width=\linewidth]{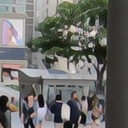} &
        \includegraphics[width=\linewidth]{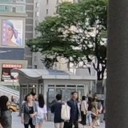} \tabularnewline

    \end{tabular}%
    \caption{Qualitative results on REDS validation set.}
    \label{fig:vsr_qualitative_reds}
\end{subfigure}
\caption{Qualitative results of video super-resolution.
Reference frames with scene name and frame number are shown on the first column, and we crop and enlarge the red box regions to show the output details of each method.
We also depict the temporal profiles of generated videos, which are represented by $x$- and $t$-axis with fixed $y$ value (blue dotted line). 
We can evaluate the restoration accuracy through the recovered images, and the smooth transition through the temporal profiles. 
By applying our method, results are better restored close to the given ground truth.}
\end{figure}
%
We first evaluate our framework on video super-resolution by  
comparing our method with several recent state-of-the-art image and video super-resolution methods.
For comparisons, we choose EnhanceNet \cite{Sajjadi_2017_ICCV} (GAN based single-image super-resolution), EDVR \cite{wang2019edvr} (pixel loss only), and TecoGAN \cite{chu2018temporally} (GAN based video super-resolution). 
%
Our method presents a learning framework for the video restoration tasks and can be applied to any generator network.
Note that we are focused on designing a better objective function rather than focusing on designing a better generator architecture. 
Therefore, we choose the generators of EDVR and TecoGAN as our generator networks (named as Ours (EDVR) and Ours (TecoGAN) respectively) because they are the most recent video super-resolution methods with the state-of-the-art performance.
By comparing ours with EDVR and TecoGAN, we verify the performance improvement when using our method over using the pixel loss and the adversarial loss respectively.

For training, we use the same training data used in EDVR and TecoGAN.
EDVR used REDS \cite{nah2019reds} train set, which contains 240 realistic video sequences having a total of 24,000 frames of $1280\times720$ resolution.
The videos have various contents including people, buildings, and other objects with variety of textures and motions. 
TecoGAN used training videos downloaded from the Internet.
The videos have diverse scenes and objects, and we use 251 sequences with a total of 30,120 frames for the training.
To compare with other methods, we set Vid4 \cite{liu2014bayesian} and REDS validation sets as the test sets, as they are commonly used for evaluating video super-resolution methods.

\begin{table}[t]
\centering
\vspace{-6pt}
\caption{Quantitative results of video super-resolution.
$\downarrow$~indicates the lower the better, and $\uparrow$ indicates the higher the better.
The best values are shown in bold and the second best values are underlined.
Our method produces perceptually better results while maintaining the temporal consistency (\ie lower LPIPS and tOF values).}
\label{tab:vsr_quantitative}
\begin{subtable}[h]{0.50\linewidth}
    \centering
    \scriptsize
    \caption{On Vid4 test set.}
    \label{tab:vsr_quantitative_vid4}
    \begin{tabular}
        {@{}p{0.34\linewidth}@{\hskip1pt}p{0.16\linewidth}@{\hskip1pt}p{0.16\linewidth}@{\hskip1pt}p{0.16\linewidth}@{\hskip1pt}p{0.16\linewidth}@{}}
        Method &
        LPIPS$\downarrow$ &
        tOF$\downarrow$ &
        PSNR$\uparrow$ &
        SSIM$\uparrow$ \\
        \hline
        
        EnhanceNet &
        0.2469 &
        0.1929 &
        20.84 &
        0.5617 \\
        
        
        
        EDVR & 
        0.2354 & 
        \textbf{0.0545} & 
        \textbf{25.86} & 
        \textbf{0.8080} \\
        
        
        Ours(EDVR) &
        \underline{0.1581} & 
        \underline{0.0568} & 
        \underline{25.13} & 
        0.7832 \\
        
        TecoGAN & 
        0.1621 & 
        0.1255 & 
        24.10 & 
        0.7588 \\
        
        Ours(TecoGAN) & 
        \textbf{0.1554} & 
        0.1248 & 
        24.55 & 
        \underline{0.7885} \\
    \end{tabular}
\end{subtable}
\hfill
\begin{subtable}[h]{0.48\linewidth}
    \centering
    \scriptsize
    \caption{On REDS validation set.}
    \label{tab:vsr_quantitative_reds}
    \begin{tabular}
        {@{}p{0.30\linewidth}@{\hskip1pt}p{0.17\linewidth}@{\hskip1pt}p{0.17\linewidth}@{\hskip1pt}p{0.17\linewidth}@{\hskip1pt}p{0.17\linewidth}@{}}
        Method &
        LPIPS$\downarrow$ &
        tOF$\downarrow$ &
        PSNR$\uparrow$ &
        SSIM$\uparrow$ \\
        \hline
        
        EDVR & 
        \underline{0.2036} & 
        \underline{3.4420} & 
        \textbf{30.88} & 
        \textbf{0.8653} \\
        
        Ours(EDVR) & 
        \textbf{0.1302} & 
        \textbf{2.8532} & 
        \underline{29.81} & 
        \underline{0.8447} \\
        
        
    \end{tabular}
\end{subtable}
\end{table}
In video super-resolution experiments, we fix the upscaling factor to 4.
We crop the ground truth frames into patches of size $128\times128$, and the size of the corresponding input patches is $32\times32$.
Based on the pretrained generator models for the pixel loss, we train $T$ only for $10^{5}$ iterations using Adam optimizer \cite{kingma2014adam} with learning rate of $10^{-5}$.
Then, we alternately train $T$ and $G$ once for another $3\times10^{4}$ iterations.
Note that pretrained $G$ is used in order to have a better initialization and avoid undesired local optima, as used in GAN based methods \cite{Ledig_2017_CVPR,Sajjadi_2017_ICCV,chu2018temporally}.

\paragraph{Comparisons}
Visual results and quantitative comparisons on Vid4 test set are shown in \fref{fig:vsr_qualitative_vid4} and \Tref{tab:vsr_quantitative_vid4} respectively.
All results of other methods are reproduced from the published codes.
We note that PSNR and SSIM values are computed on RGB channels and averaged, and we exclude the first two and the last two frames and eight border pixels to avoid errors that can occur at the borders.

EnhanceNet has the ability to hallucinate image details, but results look very noisy and fine structures are not restored well.
In addition, since the method cannot consider the temporal information, the temporal profiles look bumpy and this is observed as flickering artifacts in the videos.
Since EDVR is trained to reduce the discrepancy in the pixel space only, it has the highest PSNR and SSIM values with the lowest tOF value, but the restored images still show blurriness to some degree.





Compared to EDVR, our method shows lower PSNR and SSIM values, but has superior LPIPS value and very comparable tOF value.
Our results look visually more sharp and realistic because we further reduce the differences from the ground truth in the discriminative space.
There are some visual artifacts in the Vid4 test set, and EDVR shows slightly better tOF values compared to ours as EDVR blurs spiky pixels.

%
%
The results for the REDS validation set are in \fref{fig:vsr_qualitative_reds} and \Tref{tab:vsr_quantitative_reds}.
Here we leave out other methods that are not trained using the REDS train set, and omit temporal profiles due to the severe camera shaking.
Unlike the Vid4 test set, since there are few spiky pixels in the REDS validation set, our method has better tOF value than EDVR due to well restored sharpness and temporal consistency using our losses.

TecoGAN creates details to produce realistic videos through the adversarial loss with the VGG based perceptual loss.
TecoGAN shows low LPIPS values, but has high tOF values as the spiky pixel values in the restored details interfere with the optical flow estimation.
By applying our method to TecoGAN, we can further lower the LPIPS and the tOF values and, more surprisingly, increase PSNR and SSIM at the same time.
Existing GAN based restoration methods have focused on hallucinating arbitrary details to fool the discriminator.
In comparison, our method generates details that are closer to the given ground truth data, and achieves the better values in all four quantitative metrics than TecoGAN.
As can be seen in the Building scene, our method is able to restore fine structures that TecoGAN is unable to reconstruct.
Overall, our method shows more satisfactory results in both the sharpness and the temporal consistency than the other methods.
Please see supplementary materials including the videos for more comparisons.

%
\begin{table}[t]
\centering
\vspace{-6pt}
\caption{User study of video super-resolution and deblurring.
The value represents the preference.}
\label{tab:userstudy}
\scriptsize
\begin{subtable}[h]{0.65\linewidth}
    \centering
    \scriptsize
    \caption{Video super-resolution on Vid4 test set.}
    \label{tab:userstudy_vsr}
    \begin{tabular}
        {@{}p{0.28\linewidth}@{}p{0.14\linewidth}@{}p{0.14\linewidth}@{}p{0.14\linewidth}@{}p{0.14\linewidth}@{}p{0.14\linewidth}@{}}
        Method &
        \centering~\mbox{Calendar} &
        \centering~City &
        \centering~Foliage &
        \centering~Walk &
        \centering~Avg \tabularnewline
        \hline
        
        EDVR &
        \centering~33.3\% &
        \centering~26.7\% &
        \centering~16.7\% &
        \centering~23.3\% &
        \centering~\underline{25\%} \tabularnewline
        
        \mbox{Ours(EDVR)} &
        \centering~66.7\% &
        \centering~73.3\% &
        \centering~83.3\% &
        \centering~76.7\% &
        \centering~\textbf{75\%} \tabularnewline
        \hline
        \hline
        
        TecoGAN &
        \centering~21\% &
        \centering~69.5\% &
        \centering~23.8\% &
        \centering~42.5\% &
        \centering~\underline{39.2\%} \tabularnewline
        
        \mbox{Ours(TecoGAN)} &
        \centering~79\% &
        \centering~30.5\% &
        \centering~76.2\% &
        \centering~57.5\% &
        \centering~\textbf{60.8\%} \tabularnewline
    \end{tabular}
\end{subtable}
\hfill
\begin{subtable}[h]{0.28\linewidth}
    \centering
    \scriptsize
    \caption{Video deblurring.}
    \label{tab:userstudy_deblur}
    \begin{tabular}
        {@{}p{0.6\linewidth}@{}p{0.38\linewidth}@{}}
        Method &
        \centering~Avg \tabularnewline
        \hline
        
        OVD &
        \centering~\underline{30.7\%} \tabularnewline
        
        \mbox{Ours(OVD)} &
        \centering~\textbf{69.3\%} \tabularnewline
        \hline
        \hline
        
        STFAN &
        \centering~\underline{39.8\%} \tabularnewline
        
        \mbox{Ours(STFAN)} &
        \centering~\textbf{60.2\%} \tabularnewline
    \end{tabular}
\end{subtable}
\end{table}
\paragraph{User study}
To make sure that our method meets the human visual perception, we conducted user studies on the Vid4 test set through Amazon Mechanical Turk (\Tref{tab:userstudy_vsr}).
In the user study, we ask the users to choose more visually pleasing results between the two results, focusing on the realism with less annoying artifacts.
For the user study, two comparisons were conducted: EDVR versus Ours (EDVR), and TecoGAN versus Ours (TecoGAN).
For EDVR comparison, we randomly show two videos side by side, and ask the users to choose more pleasing one.
The comparison is performed 30 times for each scene.
For the TecoGAN comparison, we focus more on evaluating the ability to restore natural details.
To do this, we provide an user interface to easily switch between the frames from the two videos and ask the users to select the preference.
The comparison is performed 30 times for 3 uniformly sampled frames from each scene.
We exclude responses that did not click the button or had too short elapsed time for confidence.
Our method is preferred over EDVR as overall improved sharpness makes the videos look clearer.
Our results are also preferred over TecoGAN results, because our results have clean details with less artifacts.

\paragraph{Ablation study}
\Tref{tab:ablation_vsr} shows the ablation study to learn the contributions of different loss terms as well as the new fake sample setting.
We choose Ours (EDVR) as the reference model.
The pretrained EDVR model using the pixel loss achieves the best value in PSNR as it is trained to reduce the difference in the pixel space.
Only with the VGG based perceptual loss, there are not much difference from the pretrained model.
To better verify the effectiveness of our losses compared to the adversarial loss, we set our $T$ network as $D$ for adversarial training.
The adversarial loss can achieve lower LPIPS value, but tOF, PSNR, and SSIM values are much worse than our method as restored details are not much correlated with the ground truth.
Without $\mathcal{L}_{C}$, results lack high-frequency details as there is no strong guidance to reduce the content discrepancy in the discriminative space.
Adding $\mathcal{L}_{C}$ provides better LPIPS values.
We achieve the best perceptual performance while maintaining the temporal consistency using our full losses and our new fake sample setting.
Without the new fake setting, LPIPS and tOF values increase because the features learned in the loss network have less discriminating power, and consequently the generator cannot be trained to produce accurate results.

%
    
    
    
    
    
\begin{table}[t]
\centering
\vspace{-6pt}
\caption{Ablation study on Vid4 test set using the Ours (EDVR) model.}
\vspace{6pt}
\label{tab:ablation_vsr}
\begin{subtable}[h]{0.54\linewidth}
    \centering
    \scriptsize
    \begin{tabular}
        {@{}p{0.44\linewidth}@{\hskip1pt}p{0.14\linewidth}@{\hskip1pt}p{0.14\linewidth}@{\hskip1pt}p{0.14\linewidth}@{\hskip1pt}p{0.14\linewidth}@{}}
        Method &
        LPIPS$\downarrow$ &
        tOF$\downarrow$ &
        PSNR$\uparrow$ &
        SSIM$\uparrow$ \\
        \hline
        
        Pretrained &
        0.2354 & 
        \textbf{0.0545} & 
        \textbf{25.86} & 
        \textbf{0.8080} \\
        
        \mbox{Pretrained w/ VGG} &
        0.2272 & 
        0.0568 & 
        \underline{25.84} & 
        \underline{0.8073} \\
        
        \mbox{Our T w/ Adv+VGG} &
        0.1730 & 
        0.0615 & 
        23.64 & 
        0.7267 \\
        
        Ours w/o $\mathcal{L}_{C}$ &
        0.2292 & 
        \underline{0.0564} & 
        25.80 & 
        0.8050 \\
        
        Ours w/o $\mathcal{L}_{R}$ &
        0.1613 & 
        0.0576 & 
        25.16 & 
        0.7845 \\
        
        Ours w/o new fake &
        0.1876 & 
        0.0589 & 
        25.57 & 
        0.7972 \\
        
        Ours w/o init $T$ &
        \underline{0.1597} & 
        0.0578 & 
        24.99 & 
        0.7768 \\
        
        Ours &
        \textbf{0.1581} & 
        0.0568 & 
        25.13 & 
        0.7832 \\
    \end{tabular}
\end{subtable}
\hfill
\begin{subtable}[h]{0.4\linewidth}
    \centering
    \label{tab:ablation_qualitative_vid4}
    \scriptsize
    \begin{tabular}
        {@{}p{0.33\linewidth}@{\hskip1pt}p{0.33\linewidth}@{\hskip1pt}p{0.33\linewidth}@{}}
        \centering~Pretrained &
        \centering~Adv+VGG &
        \centering~w/o $\mathcal{L}_{C}$ \tabularnewline
        
        \includegraphics[width=\linewidth]{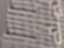} &
        \includegraphics[width=\linewidth]{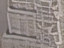} &
        \includegraphics[width=\linewidth]{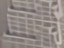} \tabularnewline
        
        \centering~w/o $\mathcal{L}_{R}$ &
        \centering~\mbox{w/o new fake} &
        \centering~{Ours} \tabularnewline
        
        \includegraphics[width=\linewidth]{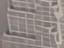} &
        \includegraphics[width=\linewidth]{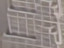} &
        \includegraphics[width=\linewidth]{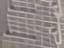} \tabularnewline
    \end{tabular}
\end{subtable}
\end{table}
%

        
        
        
        

\subsection{Video deblurring}
The goal of video deblurring is to estimate sharp video frames from the given blurry frames, caused by the camera shakes or fast moving objects.
We compare our results with DVD \cite{su2017deep}, OVD \cite{hyun2017online}, EDVR \cite{wang2019edvr}, all of which are trained to minimize the pixel loss, and STFAN \cite{zhou2019spatio} which is trained using the VGG based perceptual loss.
Here, we choose the generators of OVD, EDVR, and STFAN as our generator networks and name them as Ours (OVD), Ours (EDVR), and Ours (STFAN) respectively.
For Ours (OVD) and Ours (STFAN), we use the train set of DVD \cite{su2017deep} that includes 61 scenes with 5,708 ground truth and blurry frame pairs, with most frames having $1280\times720$ resolution.
For Ours (EDVR), we use the REDS \cite{nah2019reds} train set.
For the training, we crop the train data to $128\times128$ size patches.
The overall experimental process is similar to the video super-resolution experiment.

\begin{figure}[t]
\centering
\scriptsize
\begin{tabular}
    {@{}p{0.17\linewidth}@{\hskip2pt}}
    \tiny~\\
    \centering\mbox{\tiny~IMG\_0032/11} \\
    \includegraphics[width=\linewidth]{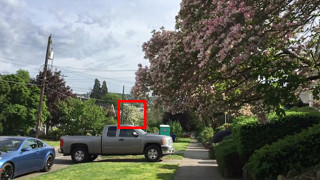} \\
    \tiny~\\
    \centering\mbox{\tiny~IMG\_0033/89} \\
    \includegraphics[width=\linewidth]{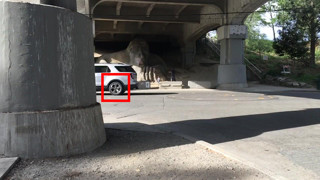} \\
\end{tabular}%
\begin{tabular}
    {@{}p{0.13\linewidth}@{\hskip2pt}p{0.13\linewidth}@{\hskip2pt}p{0.13\linewidth}@{\hskip2pt}p{0.13\linewidth}@{\hskip2pt}p{0.13\linewidth}@{\hskip2pt}p{0.13\linewidth}@{}}
    \centering\scriptsize~\mbox{DVD} &
    \centering\scriptsize~\mbox{OVD} &
    \centering\scriptsize~\mbox{Ours(OVD)} &
    \centering\scriptsize~\mbox{STFAN} &
    \centering\scriptsize~\mbox{Ours(STFAN)} &
    \centering\scriptsize~\mbox{GT} \tabularnewline
    
    \includegraphics[width=\linewidth]{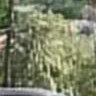} &
    \includegraphics[width=\linewidth]{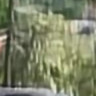} &
    \includegraphics[width=\linewidth]{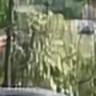} &
    \includegraphics[width=\linewidth]{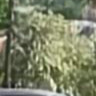} &
    \includegraphics[width=\linewidth]{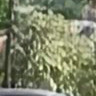} &
    \includegraphics[width=\linewidth]{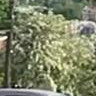} \tabularnewline
    
    \includegraphics[width=\linewidth]{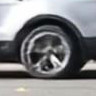} &
    \includegraphics[width=\linewidth]{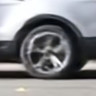} &
    \includegraphics[width=\linewidth]{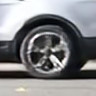} &
    \includegraphics[width=\linewidth]{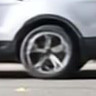} &
    \includegraphics[width=\linewidth]{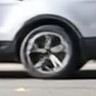} &
    \includegraphics[width=\linewidth]{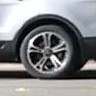} \tabularnewline
\end{tabular}%
\caption{Qualitative results of video deblurring on DVD's test set.
Applying our method increases the sharpness.}
\label{fig:deblur_qualitative}
\end{figure}

\paragraph{Comparisons}
Visual comparisons on the test set of DVD are shown in \fref{fig:deblur_qualitative}.
We reproduce OVD's results by finetuning to the DVD's train data since OVD was not trained on them.
The results of other methods are reproduced from the published codes.
In the IMG\_0032 scene, OVD has difficulty in recovering the texture of bushes as it only minimizes the pixel loss.
STFAN shows better restoration ability, because it uses an effective generator structure with the perceptual loss.
By applying our framework, the sharpness of the texture increases and the resulting images become closer to the ground truth in both cases.
Similarly, in the IMG\_0033 scene, the generator learned through our method shows better restoration ability than other methods that have difficulty in restoring a car wheel in fast motion.

Quantitative comparisons on DVD's test set and REDS validation set are shown in \Tref{tab:deblur_quantitative}.
By applying our method, we can get better LPIPS and tOF values due to better perceptually restored results with temporal consistency.
In addition, on a user study for the video deblurring, our method has 69.3\% preference over OVD and 60.2\% preference over STFAN (\Tref{tab:userstudy_deblur}).
Please see the supplementary materials for more results and user study details.
%
\begin{table}[t]
\centering
\vspace{-6pt}
\caption{Quantitative results of video deblurring.
LPIPS and tOF values are further decreased by applying our method.}
\label{tab:deblur_quantitative}
\begin{subtable}[h]{0.5\linewidth}
    \centering
    \caption{On DVD's test set.}
    \label{tab:deblur_quantitative_dvd}
    \scriptsize
    \begin{tabular}
        {@{}p{0.30\linewidth}@{\hskip1pt}p{0.17\linewidth}@{\hskip1pt}p{0.17\linewidth}@{\hskip1pt}p{0.17\linewidth}@{\hskip1pt}p{0.17\linewidth}@{}}
        Method &
        LPIPS$\downarrow$ &
        tOF$\downarrow$ &
        PSNR$\uparrow$ &
        SSIM$\uparrow$ \\
        \hline
        
        DVD &
        0.1247 & 
        5.7881 & 
        29.86 & 
        0.8809 \\ 
        
        OVD &
        0.1773 & 
        9.0234 & 
        29.95 & 
        0.8691 \\
        
        Ours(OVD) &
        0.1401 & 
        7.9667 & 
        29.36 & 
        0.8651 \\ 
        
        STFAN &
        \underline{0.1064} & 
        \underline{4.8212} & 
        \textbf{31.24} & 
        \textbf{0.9057} \\
        
        Ours(STFAN) &
        \textbf{0.0936} & 
        \textbf{4.6876} & 
        \underline{31.09} & 
        \underline{0.9049} \\ 
    \end{tabular}
\end{subtable}
\hfill
\begin{subtable}[h]{0.48\linewidth}
    \centering
    \caption{On REDS validation set.}
    \label{tab:deblur_quantitative_reds}
    \scriptsize
    \begin{tabular}
        {@{}p{0.32\linewidth}@{\hskip1pt}p{0.17\linewidth}@{\hskip1pt}p{0.17\linewidth}@{\hskip1pt}p{0.16\linewidth}@{\hskip1pt}p{0.16\linewidth}@{}}
        Method &
        LPIPS$\downarrow$ &
        tOF$\downarrow$ &
        PSNR$\uparrow$ &
        SSIM$\uparrow$ \\
        \hline
        
        EDVR & 
        \underline{0.0452} & 
        \underline{5.2339} & 
        \textbf{35.73} & 
        \textbf{0.9524} \\
        
        Ours(EDVR) & 
        \textbf{0.0411} & 
        \textbf{4.2283} & 
        \underline{34.73} & 
        \underline{0.9440} \\
    \end{tabular}
\end{subtable}
\end{table}

\section{Conclusion}
In this paper, we proposed effective loss functions for realistic video restoration without using the adversarial loss and the VGG based perceptual loss for training the generator.
The proposed content matching loss and relation matching loss successfully restore visually pleasing details that correspond to the ground truth with the new fake sample setting.
Through the experiments, we show that our method achieves the state-of-the-art performance in terms of the perceptual quality (LPIPS) and the temporal consistency (tOF) for video super-resolution and deblurring.
Our method is a general learning framework in that we can use it to any video restoration problems with ground truth data.
In the future, we would like to extend our work to other video tasks such as video-to-video translation.



\newpage

\title{Supplementary Materials}

\titlerunning{Learning the Loss Functions in a Discriminative Space for Video Restoration}
%
\author{}
\institute{}
%
\authorrunning{Y. Jo, J. Kang, S. W. Oh, S. Nam, P. Vajda, and S. J. Kim.}
%
\maketitle

\renewcommand\thesection{\Alph{section}}
\section{Loss during training}
\begin{figure}
\centering
\tiny
\begin{subfigure}[b]{0.3\linewidth}
    \centering
    \includegraphics[width=\linewidth,height=2.4cm]{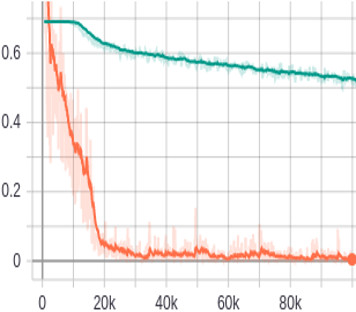}
    \caption{$\mathcal{L}_{F,C}$ ($\mathcal{L}_{D}$)}
    \label{fig:supp_loss_t}
\end{subfigure}
\hfill
\begin{subfigure}[b]{0.3\linewidth}
    \centering
    \includegraphics[width=\linewidth,height=2.4cm]{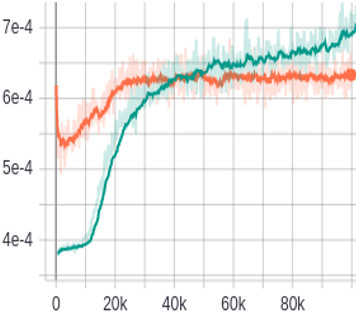}
    \caption{$\mathcal{L}_{G}$}
    \label{fig:supp_loss_g}
\end{subfigure}
\hfill
\begin{subfigure}[b]{0.3\linewidth}
    \centering
    \includegraphics[width=\linewidth,height=2.4cm]{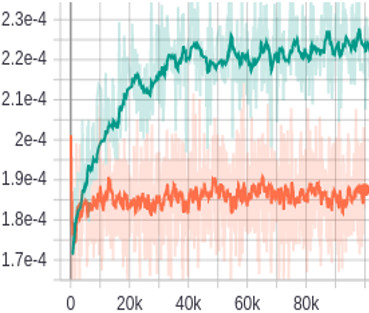}
    \caption{$\mathcal{L}_{P}$}
    \label{fig:supp_loss_p}
\end{subfigure}
\\~\\
\begin{subfigure}[b]{0.24\linewidth}
    \centering
    \includegraphics[width=\linewidth,height=2.4cm]{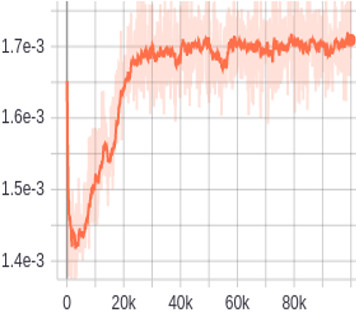}
    \caption{$\mathcal{L}_{C}$}
    \label{fig:supp_loss_c}
\end{subfigure}
\hfill
\begin{subfigure}[b]{0.24\linewidth}
    \centering
    \includegraphics[width=\linewidth,height=2.4cm]{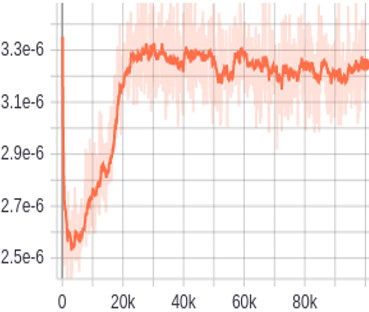}
    \caption{$\mathcal{L}_{R}$}
    \label{fig:supp_loss_r}
\end{subfigure}
\hfill
\begin{subfigure}[b]{0.24\linewidth}
    \centering
    \includegraphics[width=\linewidth,height=2.4cm]{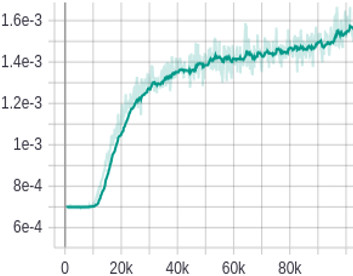}
    \caption{$\mathcal{L}_{Adv}$}
    \label{fig:supp_loss_adv}
\end{subfigure}
\hfill
\begin{subfigure}[b]{0.24\linewidth}
    \centering
    \includegraphics[width=\linewidth,height=2.4cm]{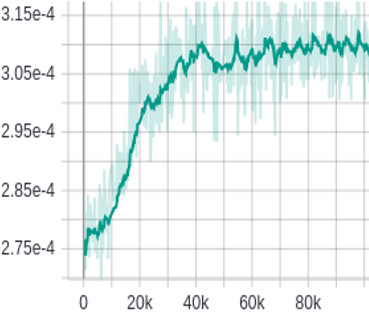}
    \caption{$\mathcal{L}_{VGG}$}
    \label{fig:supp_loss_vgg}
\end{subfigure}
\caption{Training loss comparisons with our method and the adversarial loss.
By using our method, the losses are stabilized after 30K iterations of training.}
\label{fig:supp_training_loss}
\end{figure}

We depict training losses to see how learning progresses when using our method compared to the adversarial loss.
Specifically, $T$ for our method and $D$ for the adversarial loss have same network structures.
In \fref{fig:supp_training_loss}, the losses of our method are shown in orange and the losses of the adversarial loss with the VGG perceptual loss are shown in green.
Note that our $T$ is trained by $\mathcal{L}_{F,C}$, and $G$ is trained by $\mathcal{L}_{G} = \mathcal{L}_{C} + \mathcal{L}_{R} + \mathcal{L}_{P}$.
Similarly, for the adversarial loss based model, $D$ is trained by $\mathcal{L}_{D}$, and $G$ is trained by $\mathcal{L}_{G} = \mathcal{L}_{Adv} + \mathcal{L}_{VGG} + \mathcal{L}_{P}$.
By using our method, both $\mathcal{L}_{F,C}$ and $\mathcal{L}_{G}$ are stabilized after 30K iterations of training.
However, in the adversarial training, $\mathcal{L}_{D}$ and $\mathcal{L}_{G}$ seems difficult to stabilize.
If $G$ produces realistic outputs, then $\mathcal{L}_{adv}$ should be decreased, but it actually does not.
We can see that our proposed loss functions are more effectively trains $G$ than the adversarial loss.

\section{Layer-level loss analysis}
Our loss functions use features from all layers of the feature extractor $F$.
To see what is learned from each layer, we run experiments on video super-resolution by setting the loss function to use only one of the layers.
We use Ours (EDVR) for video super-resolution as a reference model, and the results are shown in \Tref{tab:supp_vsr_layerlevelloss_quantitative}.
Early layers are closer to the pixel space, and they try to restore the exact pixel value for the exact location resulting in better PSNR and SSIM values in addition to the temporal smoothness.
Higher level layers can learn features that better understand a given task and the exact pixel location is considered less.
The higher layers tend to create high-frequency details for the video super-resolution and have better LPIPS values, but show more noise.
With all the layers, all the properties form an ensemble, providing high visual quality and the best LPIPS value and comparable tOF value.
%
\begin{table}[h]
\small
\centering
\vspace{-6pt}
\caption{Quantitative results of layer-level loss for the generator.
Full model is Ours (EDVR).
With all layers, both high accuracy and temporal consistency are achieved.}
\label{tab:supp_vsr_layerlevelloss_quantitative}
\vspace{6pt}
\begin{tabular}
    {@{}p{0.34\linewidth}@{\hskip1pt}p{0.15\linewidth}@{\hskip1pt}p{0.15\linewidth}@{\hskip1pt}p{0.15\linewidth}@{\hskip1pt}p{0.15\linewidth}@{}}
    Layer &
    LPIPS $\downarrow$ &
    tOF $\downarrow$ &
    PSNR $\uparrow$ &
    SSIM $\uparrow$ \\
    \hline
    
    First & 
    0.2287 & 
    \textbf{0.0544} & 
    \textbf{25.81} & 
    \textbf{0.8107} \\
    
    Second & 
    0.2273 & 
    \underline{0.0562} & 
    \underline{25.68} & 
    \underline{0.8082} \\
    
    Third & 
    0.1874 & 
    0.0604 & 
    25.39 & 
    0.7888 \\
    
    Last & 
    \underline{0.1841} & 
    0.0993 & 
    24.36 & 
    0.7512 \\
    
    All & 
    \textbf{0.1581} & 
    0.0568 & 
    25.13 & 
    0.7832 \\
\end{tabular}
\\\vspace{6pt}
\begin{tabular}
    {@{}p{0.15\linewidth}@{\hskip2pt}p{0.15\linewidth}@{\hskip2pt}p{0.15\linewidth}@{\hskip2pt}p{0.15\linewidth}@{\hskip2pt}p{0.15\linewidth}@{}}
    \centering\footnotesize~\mbox{First} &
    \centering\footnotesize~\mbox{Second} &
    \centering\footnotesize~\mbox{Third} &
    \centering\footnotesize~\mbox{Last} &
    \centering\footnotesize~\mbox{All} \tabularnewline
    
    \includegraphics[width=\linewidth]{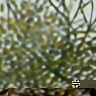} &
    \includegraphics[width=\linewidth]{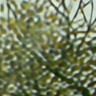} &
    \includegraphics[width=\linewidth]{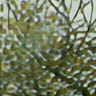} &
    \includegraphics[width=\linewidth]{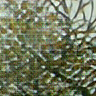} &
    \includegraphics[width=\linewidth]{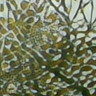} \tabularnewline
    
    \includegraphics[width=\linewidth]{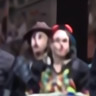} &
    \includegraphics[width=\linewidth]{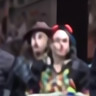} &
    \includegraphics[width=\linewidth]{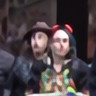} &
    \includegraphics[width=\linewidth]{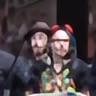} &
    \includegraphics[width=\linewidth]{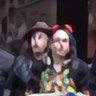} \tabularnewline
\end{tabular}%
\end{table}
%

\section{Effect of weight balancing}
We conduct additional experiments to inspect the effect of weight balancing between the two proposed loss functions $\mathcal{L}_C$ and $\mathcal{L}_R$.
Again, we use Ours (EDVR) as a reference model, and the results are shown in \Tref{tab:supp_vsr_weightbalancing_quantitative}.
Generally, the weight of the adversarial loss term in GAN based methods takes effect on the amount of hallucinated details.
However, our method gives fairly consistent results regardless of the different weight combinations.
The best result is obtained when the ratio is 1:1, but the performance gap is relatively small.
We infer that this is because both loss functions are computed on the same feature space from $F$.
In comparison, the adversarial loss is computed on a different space.
\begin{table}[h]
\small
\centering
\vspace{-6pt}
\caption{Results of different weight combinations for $\mathcal{L}_C$ and $\mathcal{L}_R$.
The best result is obtained when the ratio is 1:1, but the performance gap is not much regardless of different weight combinations.}
\label{tab:supp_vsr_weightbalancing_quantitative}
\vspace{6pt}
\begin{tabular}
    {@{}p{0.34\linewidth}@{\hskip1pt}p{0.15\linewidth}@{\hskip1pt}p{0.15\linewidth}@{\hskip1pt}p{0.15\linewidth}@{\hskip1pt}p{0.15\linewidth}@{}}
    Combination &
    LPIPS $\downarrow$ &
    tOF $\downarrow$ &
    PSNR $\uparrow$ &
    SSIM $\uparrow$ \\
    \hline
    
    $\mathcal{L}_C+\mathcal{L}_R$ & 
    \textbf{0.1581} & 
    \underline{0.0568} & 
    \textbf{25.13} & 
    \textbf{0.7832} \\
    
    $10\cdot\mathcal{L}_C+\mathcal{L}_R$ & 
    0.1606 & 
    0.0577 & 
    24.78 & 
    0.7754 \\
    
    $100\cdot\mathcal{L}_C+\mathcal{L}_R$ & 
    0.1628 & 
    0.0581 & 
    24.72 & 
    0.7760 \\
    
    $\mathcal{L}_C+10\cdot\mathcal{L}_R$ & 
    \underline{0.1603} & 
    \textbf{0.0564} & 
    24.67 & 
    0.7743 \\
    
    $\mathcal{L}_C+100\cdot\mathcal{L}_R$ & 
    0.1611 & 
    0.0576 & 
    \underline{24.92} & 
    \underline{0.7778} \\
\end{tabular}
\end{table}

\section{More visual results}
\begin{figure}
\centering
\footnotesize
\begin{tabular}
    {@{}p{0.12\linewidth}@{\hskip2pt}}
    ~\\
    ~\\
    \centering\tiny{Calendar/32} \\
    \includegraphics[width=\linewidth]{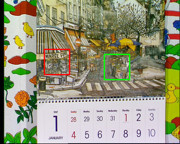} \\
    ~\\
    ~\\
    ~\\
    ~\\
    ~\\
    ~\\
    ~\\
    ~\\
    ~\\
    ~\\
    ~\\
    ~\\
    ~\\
    ~\\
    ~\\
    \centering\tiny{City/10} \\
    \includegraphics[width=\linewidth]{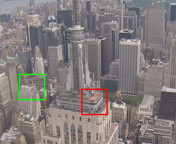} \\
    ~\\
    ~\\
    ~\\
    ~\\
    ~\\
    ~\\
    ~\\
    ~\\
    ~\\
    ~\\
    ~\\
    ~\\
    \centering\tiny{Foliage/10} \\
    \includegraphics[width=\linewidth]{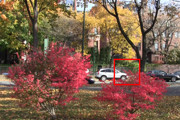} \\
    ~\\
    ~\\
    ~\\
    ~\\
    ~\\
    ~\\
    ~\\
    ~\\
    ~\\
    ~\\
    ~\\
    ~\\
    ~\\
    ~\\
    \centering\tiny{Walk/31} \\
    \includegraphics[width=\linewidth]{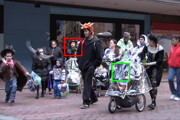} \\
    ~\\
    ~\\
    ~\\
\end{tabular}%
\begin{tabular}
    {@{}p{0.12\linewidth}@{\hskip2pt}p{0.12\linewidth}@{\hskip2pt}p{0.12\linewidth}@{\hskip2pt}p{0.12\linewidth}@{\hskip2pt}p{0.12\linewidth}@{\hskip2pt}p{0.12\linewidth}@{\hskip2pt}p{0.12\linewidth}@{}}
    \centering\scriptsize~\mbox{EnhanceNet} &
    \centering\scriptsize~\mbox{RBPN} &
    \centering\scriptsize~\mbox{EDVR} &
    \centering\scriptsize~\mbox{Ours(EDVR)} &
    \centering\scriptsize~\mbox{TecoGAN} &
    \centering\scriptsize~\mbox{Ours(TecoGAN)} &
    \centering\scriptsize~\mbox{GT} \tabularnewline
    
    \includegraphics[width=\linewidth]{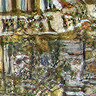} &
    \includegraphics[width=\linewidth]{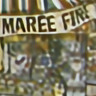} &
    \includegraphics[width=\linewidth]{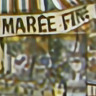} &
    \includegraphics[width=\linewidth]{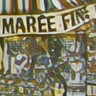} &
    \includegraphics[width=\linewidth]{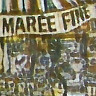} &
    \includegraphics[width=\linewidth]{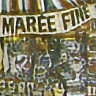} &
    \includegraphics[width=\linewidth]{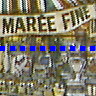} \tabularnewline
    
    \includegraphics[width=\linewidth]{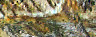} &
    \includegraphics[width=\linewidth]{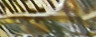} &
    \includegraphics[width=\linewidth]{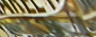} &
    \includegraphics[width=\linewidth]{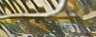} &
    \includegraphics[width=\linewidth]{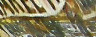} &
    \includegraphics[width=\linewidth]{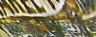} &
    \includegraphics[width=\linewidth]{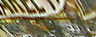} \tabularnewline
    
    \includegraphics[width=\linewidth]{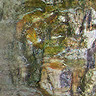} &
    \includegraphics[width=\linewidth]{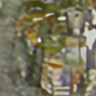} &
    \includegraphics[width=\linewidth]{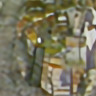} &
    \includegraphics[width=\linewidth]{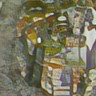} &
    \includegraphics[width=\linewidth]{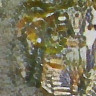} &
    \includegraphics[width=\linewidth]{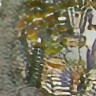} &
    \includegraphics[width=\linewidth]{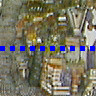} \tabularnewline
    
    \includegraphics[width=\linewidth]{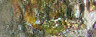} &
    \includegraphics[width=\linewidth]{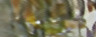} &
    \includegraphics[width=\linewidth]{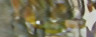} &
    \includegraphics[width=\linewidth]{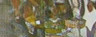} &
    \includegraphics[width=\linewidth]{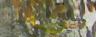} &
    \includegraphics[width=\linewidth]{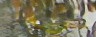} &
    \includegraphics[width=\linewidth]{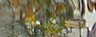} \tabularnewline
    ~\\
    
    \includegraphics[width=\linewidth]{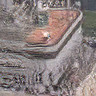} &
    \includegraphics[width=\linewidth]{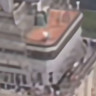} &
    \includegraphics[width=\linewidth]{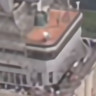} &
    \includegraphics[width=\linewidth]{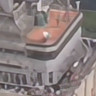} &
    \includegraphics[width=\linewidth]{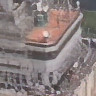} &
    \includegraphics[width=\linewidth]{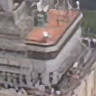} &
    \includegraphics[width=\linewidth]{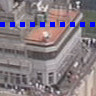} \\
    
    \includegraphics[width=\linewidth]{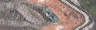} &
    \includegraphics[width=\linewidth]{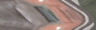} &
    \includegraphics[width=\linewidth]{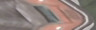} &
    \includegraphics[width=\linewidth]{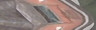} &
    \includegraphics[width=\linewidth]{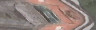} &
    \includegraphics[width=\linewidth]{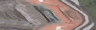} &
    \includegraphics[width=\linewidth]{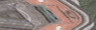} \\
    
    \includegraphics[width=\linewidth]{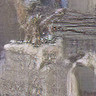} &
    \includegraphics[width=\linewidth]{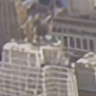} &
    \includegraphics[width=\linewidth]{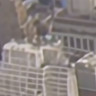} &
    \includegraphics[width=\linewidth]{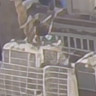} &
    \includegraphics[width=\linewidth]{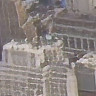} &
    \includegraphics[width=\linewidth]{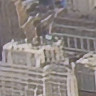} &
    \includegraphics[width=\linewidth]{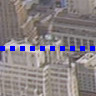} \\
    
    \includegraphics[width=\linewidth]{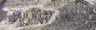} &
    \includegraphics[width=\linewidth]{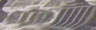} &
    \includegraphics[width=\linewidth]{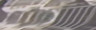} &
    \includegraphics[width=\linewidth]{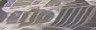} &
    \includegraphics[width=\linewidth]{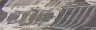} &
    \includegraphics[width=\linewidth]{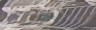} &
    \includegraphics[width=\linewidth]{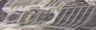} \\
    ~\\
    
    \includegraphics[width=\linewidth]{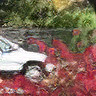} &
    \includegraphics[width=\linewidth]{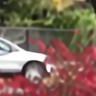} &
    \includegraphics[width=\linewidth]{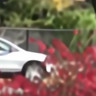} &
    \includegraphics[width=\linewidth]{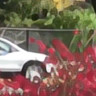} &
    \includegraphics[width=\linewidth]{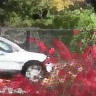} &
    \includegraphics[width=\linewidth]{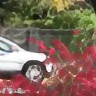} &
    \includegraphics[width=\linewidth]{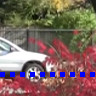} \\
    
    \includegraphics[width=\linewidth]{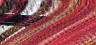} &
    \includegraphics[width=\linewidth]{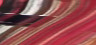} &
    \includegraphics[width=\linewidth]{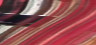} &
    \includegraphics[width=\linewidth]{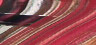} &
    \includegraphics[width=\linewidth]{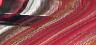} &
    \includegraphics[width=\linewidth]{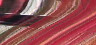} &
    \includegraphics[width=\linewidth]{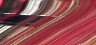} \\

    ~\\
    
    \includegraphics[width=\linewidth]{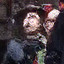} &
    \includegraphics[width=\linewidth]{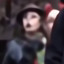} &
    \includegraphics[width=\linewidth]{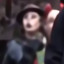} &
    \includegraphics[width=\linewidth]{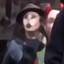} &
    \includegraphics[width=\linewidth]{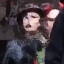} &
    \includegraphics[width=\linewidth]{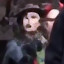} &
    \includegraphics[width=\linewidth]{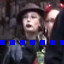} \\
    
    \includegraphics[width=\linewidth]{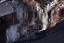} &
    \includegraphics[width=\linewidth]{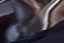} &
    \includegraphics[width=\linewidth]{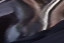} &
    \includegraphics[width=\linewidth]{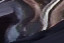} &
    \includegraphics[width=\linewidth]{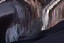} &
    \includegraphics[width=\linewidth]{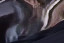} &
    \includegraphics[width=\linewidth]{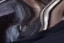} \\
    
    \includegraphics[width=\linewidth]{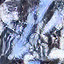} &
    \includegraphics[width=\linewidth]{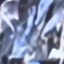} &
    \includegraphics[width=\linewidth]{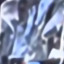} &
    \includegraphics[width=\linewidth]{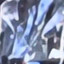} &
    \includegraphics[width=\linewidth]{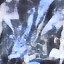} &
    \includegraphics[width=\linewidth]{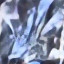} &
    \includegraphics[width=\linewidth]{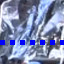} \\
    
    \includegraphics[width=\linewidth]{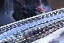} &
    \includegraphics[width=\linewidth]{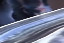} &
    \includegraphics[width=\linewidth]{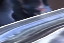} &
    \includegraphics[width=\linewidth]{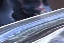} &
    \includegraphics[width=\linewidth]{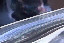} &
    \includegraphics[width=\linewidth]{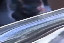} &
    \includegraphics[width=\linewidth]{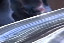} \\
\end{tabular}%
\caption{Qualitative results of video super-resolution on Vid4 test set.}
\label{fig:supp_vsr_qualitative_vid4}
\end{figure}

\begin{figure}
\centering
\tiny
\begin{tabular}
    {@{}p{0.20\linewidth}@{}}
    ~\\
    ~\\
    ~\\
    \centering 000/11 \\
    \includegraphics[width=\linewidth]{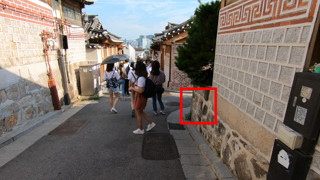} \\
    ~\\
    ~\\
    ~\\
    ~\\
    ~\\
    ~\\
    \centering 001/60 \\
    \includegraphics[width=\linewidth]{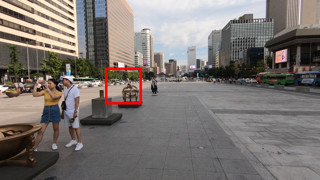} \\
    ~\\
    ~\\
    ~\\
    ~\\
    ~\\
    ~\\
    \centering 009/77 \\
    \includegraphics[width=\linewidth]{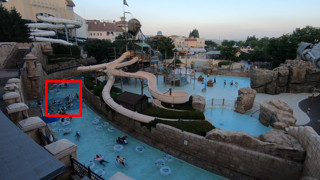} \\
    ~\\
    ~\\
    ~\\
    ~\\
    ~\\
    ~\\
    \centering 011/30 \\
    \includegraphics[width=\linewidth]{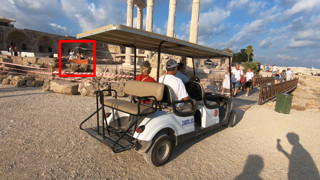} \\
    ~\\
    ~\\
    ~\\
    ~\\
    ~\\
    ~\\
    \centering 014/41 \\
    \includegraphics[width=\linewidth]{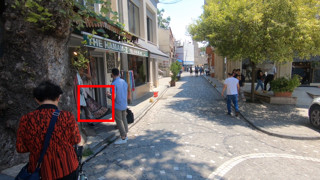} \\
    ~\\
    ~\\
    ~\\
    ~\\
    ~\\
    ~\\
    \centering 015/55 \\
    \includegraphics[width=\linewidth]{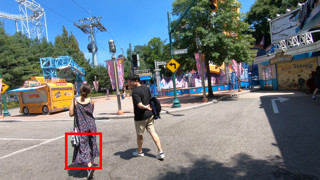} \\
    ~\\
\end{tabular}%
\footnotesize
\begin{tabular}
    {@{\hskip2pt}p{0.20\linewidth}@{\hskip2pt}p{0.20\linewidth}@{\hskip2pt}p{0.20\linewidth}@{}}
    \centering~\scriptsize\scriptsize\mbox{EDVR} &
    \centering~\scriptsize\mbox{Ours(EDVR)} &
    \centering~\scriptsize\mbox{GT} \tabularnewline
    
    \includegraphics[width=\linewidth]{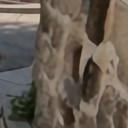} &
    \includegraphics[width=\linewidth]{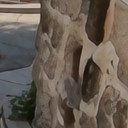} &
    \includegraphics[width=\linewidth]{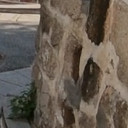} \tabularnewline
    ~\\
    
    \includegraphics[width=\linewidth]{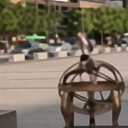} &
    \includegraphics[width=\linewidth]{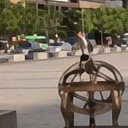} &
    \includegraphics[width=\linewidth]{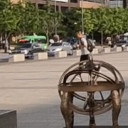} \tabularnewline
    ~\\
    
    \includegraphics[width=\linewidth]{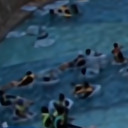} &
    \includegraphics[width=\linewidth]{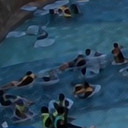} &
    \includegraphics[width=\linewidth]{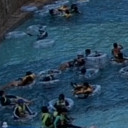} \tabularnewline
    ~\\
    
    \includegraphics[width=\linewidth]{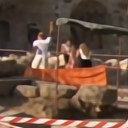} &
    \includegraphics[width=\linewidth]{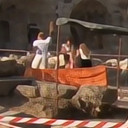} &
    \includegraphics[width=\linewidth]{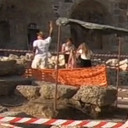} \tabularnewline
    ~\\
    
    \includegraphics[width=\linewidth]{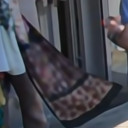} &
    \includegraphics[width=\linewidth]{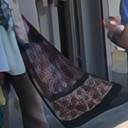} &
    \includegraphics[width=\linewidth]{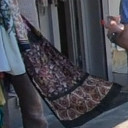} \tabularnewline
    ~\\
    
    \includegraphics[width=\linewidth]{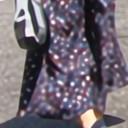} &
    \includegraphics[width=\linewidth]{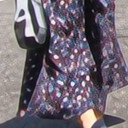} &
    \includegraphics[width=\linewidth]{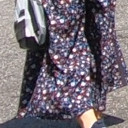} \tabularnewline
\end{tabular}%
\caption{Qualitative results of video super-resolution on REDS validation set.}
\label{fig:supp_vsr_qualitative_reds}
\end{figure}

\begin{figure}
\centering
\tiny
\begin{tabular}
    {@{}p{0.17\linewidth}@{\hskip2pt}}
    ~\\
    ~\\
    \centering\mbox{\tiny~720p\_240fps\_2/88} \\
    \includegraphics[width=\linewidth]{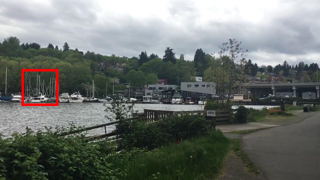} \\
    ~\\
    ~\\
    ~\\
    \centering\mbox{\tiny~IMG\_0003/69} \\
    \includegraphics[width=\linewidth]{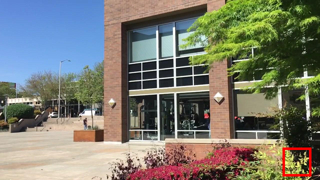} \\
    ~\\
    ~\\
    ~\\
    \centering\mbox{\tiny~IMG\_0021/54} \\
    \includegraphics[width=\linewidth]{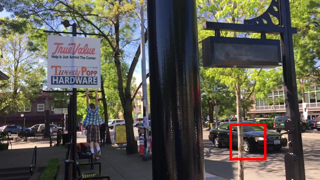} \\
    ~\\
    ~\\
    ~\\
    \centering\mbox{\tiny~IMG\_0031/76} \\
    \includegraphics[width=\linewidth]{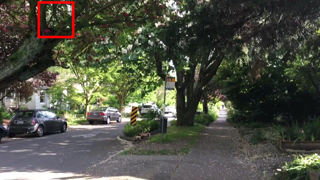} \\
    ~\\
    ~\\
    ~\\
    \centering\mbox{\tiny~IMG\_0032/89} \\
    \includegraphics[width=\linewidth]{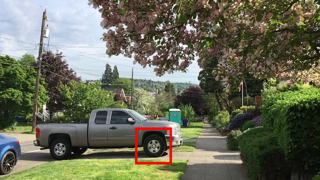} \\
    ~\\
    ~\\
    ~\\
    \centering\mbox{\tiny~IMG\_0037/76} \\
    \includegraphics[width=\linewidth]{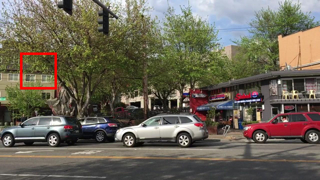} \\
    ~\\
\end{tabular}%
\footnotesize
\begin{tabular}
    {@{}p{0.13\linewidth}@{\hskip2pt}p{0.13\linewidth}@{\hskip2pt}p{0.13\linewidth}@{\hskip2pt}p{0.13\linewidth}@{\hskip2pt}p{0.13\linewidth}@{\hskip2pt}p{0.13\linewidth}@{}}
    \centering\scriptsize~\mbox{DVD} &
    \centering\scriptsize~\mbox{OVD} &
    \centering\scriptsize~\mbox{Ours(OVD)} &
    \centering\scriptsize~\mbox{STFAN} &
    \centering\scriptsize~\mbox{Ours(STFAN)} &
    \centering\scriptsize~\mbox{GT} \tabularnewline
    
    \includegraphics[width=\linewidth]{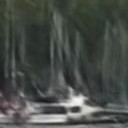} &
    \includegraphics[width=\linewidth]{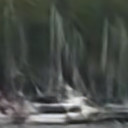} &
    \includegraphics[width=\linewidth]{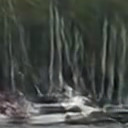} &
    \includegraphics[width=\linewidth]{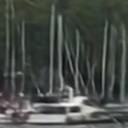} &
    \includegraphics[width=\linewidth]{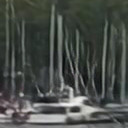} &
    \includegraphics[width=\linewidth]{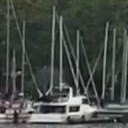} \\
    ~\\
    
    \includegraphics[width=\linewidth]{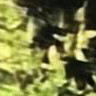} &
    \includegraphics[width=\linewidth]{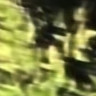} &
    \includegraphics[width=\linewidth]{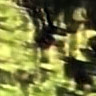} &
    \includegraphics[width=\linewidth]{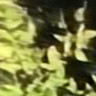} &
    \includegraphics[width=\linewidth]{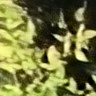} &
    \includegraphics[width=\linewidth]{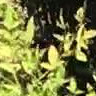} \\
    ~\\
    
    \includegraphics[width=\linewidth]{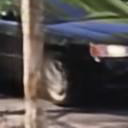} &
    \includegraphics[width=\linewidth]{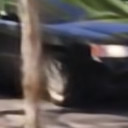} &
    \includegraphics[width=\linewidth]{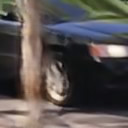} &
    \includegraphics[width=\linewidth]{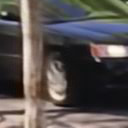} &
    \includegraphics[width=\linewidth]{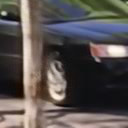} &
    \includegraphics[width=\linewidth]{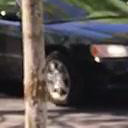} \\
    ~\\
    
    \includegraphics[width=\linewidth]{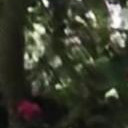} &
    \includegraphics[width=\linewidth]{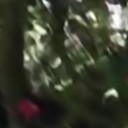} &
    \includegraphics[width=\linewidth]{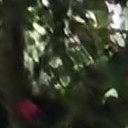} &
    \includegraphics[width=\linewidth]{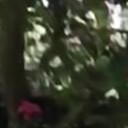} &
    \includegraphics[width=\linewidth]{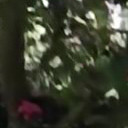} &
    \includegraphics[width=\linewidth]{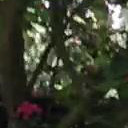} \\
    ~\\
    
    \includegraphics[width=\linewidth]{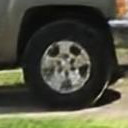} &
    \includegraphics[width=\linewidth]{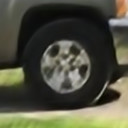} &
    \includegraphics[width=\linewidth]{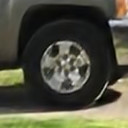} &
    \includegraphics[width=\linewidth]{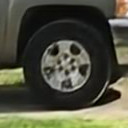} &
    \includegraphics[width=\linewidth]{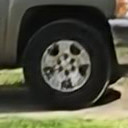} &
    \includegraphics[width=\linewidth]{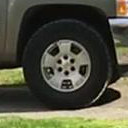} \\
    ~\\
    
    \includegraphics[width=\linewidth]{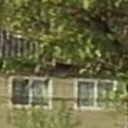} &
    \includegraphics[width=\linewidth]{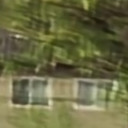} &
    \includegraphics[width=\linewidth]{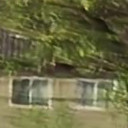} &
    \includegraphics[width=\linewidth]{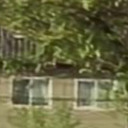} &
    \includegraphics[width=\linewidth]{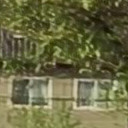} &
    \includegraphics[width=\linewidth]{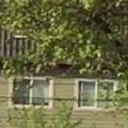} \\
\end{tabular}%
\caption{Qualitative results of video deblurring on the quantitative test set of DVD.
Best viewed in zoom.}
\label{fig:supp_deblur_qualitative_dvd}
\end{figure}

\begin{figure}
\centering
\tiny
\begin{tabular}
    {@{}p{0.20\linewidth}@{\hskip2pt}}
    ~\\
    ~\\
    ~\\
    \centering\mbox{\tiny~Anita/40} \\
    \includegraphics[width=\linewidth]{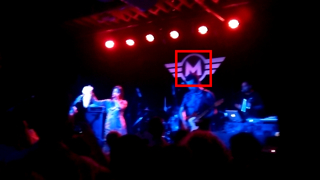} \\
    ~\\
    ~\\
    ~\\
    \centering\mbox{\tiny~Cutting/71} \\
    \includegraphics[width=\linewidth]{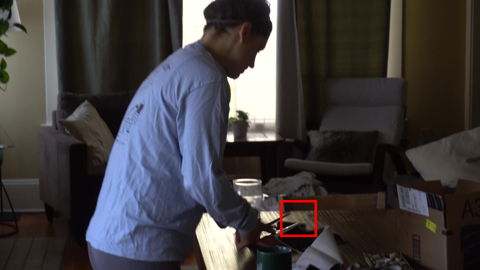} \\
    ~\\
    ~\\
    ~\\
    \centering\mbox{\tiny~Street2/31} \\
    \includegraphics[width=\linewidth]{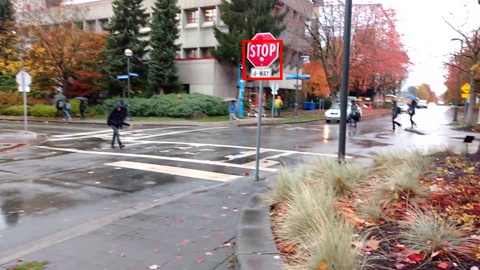} \\
    ~\\
    ~\\
    ~\\
    \centering\mbox{\tiny~Sup/41} \\
    \includegraphics[width=\linewidth]{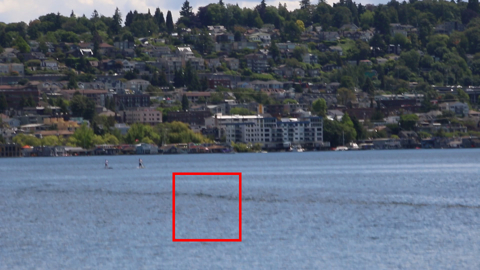} \\
    ~\\
\end{tabular}%
\footnotesize
\begin{tabular}
    {@{}p{0.15\linewidth}@{\hskip2pt}p{0.15\linewidth}@{\hskip2pt}p{0.15\linewidth}@{\hskip2pt}p{0.15\linewidth}@{\hskip2pt}p{0.15\linewidth}@{}}
    \centering\scriptsize~\mbox{DVD} &
    \centering\scriptsize~\mbox{OVD} &
    \centering\scriptsize~\mbox{Ours(OVD)} &
    \centering\scriptsize~\mbox{STFAN} &
    \centering\scriptsize~\mbox{Ours(STFAN)} \tabularnewline
    
    \includegraphics[width=\linewidth]{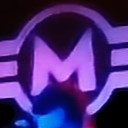} &
    \includegraphics[width=\linewidth]{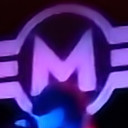} &
    \includegraphics[width=\linewidth]{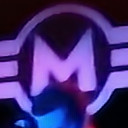} &
    \includegraphics[width=\linewidth]{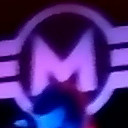} &
    \includegraphics[width=\linewidth]{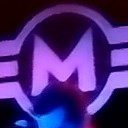} \\
    ~\\
    
    
    \includegraphics[width=\linewidth]{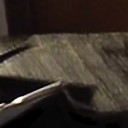} &
    \includegraphics[width=\linewidth]{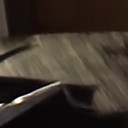} &
    \includegraphics[width=\linewidth]{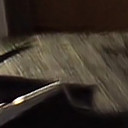} &
    \includegraphics[width=\linewidth]{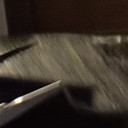} &
    \includegraphics[width=\linewidth]{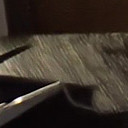} \\
    ~\\
    
    \includegraphics[width=\linewidth]{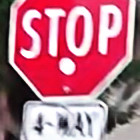} &
    \includegraphics[width=\linewidth]{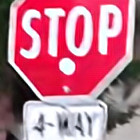} &
    \includegraphics[width=\linewidth]{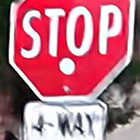} &
    \includegraphics[width=\linewidth]{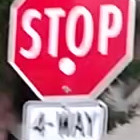} &
    \includegraphics[width=\linewidth]{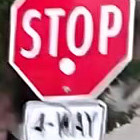} \\
    ~\\
    
    \includegraphics[width=\linewidth]{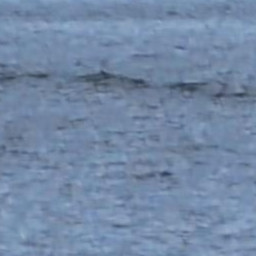} &
    \includegraphics[width=\linewidth]{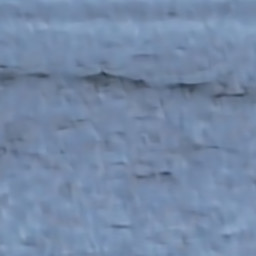} &
    \includegraphics[width=\linewidth]{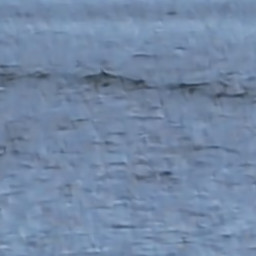} &
    \includegraphics[width=\linewidth]{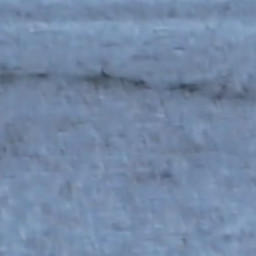} &
    \includegraphics[width=\linewidth]{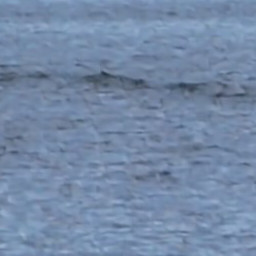} \\
\end{tabular}%
\caption{Qualitative results of video deblurring on the qualitative test set of DVD.
Our method tries to restore the results to look sharper, therefore, wood and water textures are well recovered in the Cutting and Sup scenes respectively.
}
\label{fig:supp_deblur_qualitative_dvd_real}
\end{figure}

More visual results of video super-resolution on Vid4 test set and REDS validation set are shown in \fref{fig:supp_vsr_qualitative_vid4} and \fref{fig:supp_vsr_qualitative_reds} respectively.
We additionally show the results of RBPN \cite{RBPN2019} for Vid4.
Also, more visual results of video deblurring on the test sets of DVD are shown in \fref{fig:supp_deblur_qualitative_dvd} and \fref{fig:supp_deblur_qualitative_dvd_real}.
We have also included video demos in the supplmentary video.

\section{User study on video deblurring}
\begin{figure}
\centering
\small
\includegraphics[width=\linewidth]{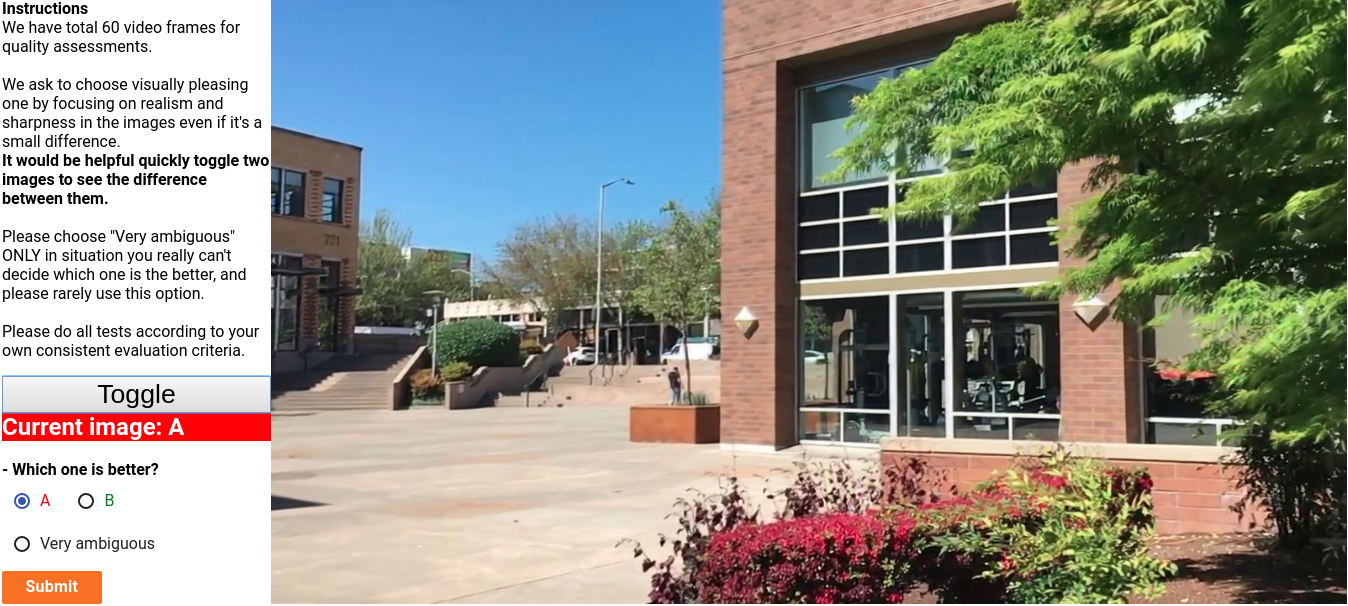}
\caption{Our user interface for the user study.
Two video frames from different methods are switchable by clicking the toggle button.}
\label{fig:supp_deblur_userstudy_ui}
\end{figure}

We conduct a user study on the results of video deblurring on the test set of DVD \cite{su2017deep}.
In the user study, following two comparisons are conducted to see the improvements when using our method: OVD \cite{hyun2017online} versus Ours (OVD), and STFAN \cite{zhou2019spatio} versus Ours (STFAN).
As comparing two videos is difficult due to camera shaking, we provide an user interface to easily switch and view between two video frames, and ask the raters to choose the more preferred ones (\fref{fig:supp_deblur_userstudy_ui}).
The comparison is performed 30 times for uniformly sampled 3 frames from each scene.
The user study is conducted with Amazon Mechanical Turk, and the results of this study are shown in \Tref{tab:supp_deblur_userstudy}.
The preference of each scene is calculated by averaging the preference of 3 frames.
In both comparisons, people prefer our method on average.

%
\begin{table}[h]
\small
\centering
\vspace{-6pt}
\caption{User study of video deblurring on the quantitative test set of DVD.
It shows the preference of each scene and total average in percent.
People prefer our method on average.}
\label{tab:supp_deblur_userstudy}
\vspace{6pt}
\begin{tabular}
    {@{}p{0.18\linewidth}@{}p{0.08\linewidth}@{}p{0.08\linewidth}@{}p{0.07\linewidth}@{}p{0.07\linewidth}@{}p{0.07\linewidth}@{}p{0.07\linewidth}@{}p{0.07\linewidth}@{}p{0.07\linewidth}@{}p{0.07\linewidth}@{}p{0.07\linewidth}@{}p{0.08\linewidth}@{}}
    Method &
    \centering~{720p} &
    \centering~0003 &
    \centering~0021 &
    \centering~0030 &
    \centering~0031 &
    \centering~0032 &
    \centering~0033 &
    \centering~0037 &
    \centering~0039 &
    \centering~0049 &
    \centering~Avg \tabularnewline
    \hline
    
    OVD &
    \centering~28.4 &
    \centering~17.0 &
    \centering~42.5 &
    \centering~34.8 &
    \centering~26.7 &
    \centering~38.4 &
    \centering~21.1 &
    \centering~17.6 &
    \centering~45.3 &
    \centering~34.9 &
    \centering~\underline{30.7} \tabularnewline
    
    Ours(OVD) &
    \centering~71.6 &
    \centering~83.0 &
    \centering~57.5 &
    \centering~65.2 &
    \centering~73.3 &
    \centering~61.6 &
    \centering~78.9 &
    \centering~82.4 &
    \centering~54.7 &
    \centering~65.1 &
    \centering~\textbf{69.3} \tabularnewline
    \hline
    \hline
    
    STFAN &
    \centering~41.8 &
    \centering~34.2 &
    \centering~35.4 &
    \centering~59.7 &
    \centering~41.4 &
    \centering~19.3 &
    \centering~31.6 &
    \centering~39.1 &
    \centering~40.0 &
    \centering~55.0 &
    \centering~\underline{39.8} \tabularnewline
    
    Ours(STFAN) &
    \centering~58.2 &
    \centering~65.8 &
    \centering~64.6 &
    \centering~40.3 &
    \centering~58.6 &
    \centering~80.7 &
    \centering~68.4 &
    \centering~60.9 &
    \centering~60.0 &
    \centering~45.0 &
    \centering~\textbf{60.2} \tabularnewline
\end{tabular}
\end{table}
%

    
    
    
    

%

%

%
%
\clearpage
\bibliographystyle{splncs04}
\bibliography{manuscript}
\end{document}